\documentclass[twocolumn]{aastex63}
\usepackage{diagbox}
\usepackage{multirow}

\received{----}
\revised{----}
\accepted{----}

\begin{document}

\title{On the true fractions of repeating and non-repeating FRB sources}

\author{Shunke Ai}
\affiliation{Department of Physics and Astronomy, University of Nevada Las Vegas, Las Vegas, NV 89154, USA \\ ais1@unlv.nevada.edu, zhang@physics.unlv.edu}

\author{He Gao}
\affiliation{Department of Astronomy, Beijing Normal University, Beijing 100875, People's Republic of China; gaohe@bnu.edu.cn}

\author{Bing Zhang}
\affiliation{Department of Physics and Astronomy, University of Nevada Las Vegas, Las Vegas, NV 89154, USA \\ ais1@unlv.nevada.edu, zhang@physics.unlv.edu}

\begin{abstract}
Observationally, fast radio bursts (FRBs) can be divided into repeating and apparently non-repeating (one-off) ones. It is unclear whether all FRBs repeat and whether there are genuine non-repeating FRBs. We attempt to address these questions using Monte Carlo simulations. We define a parameter $T_c$ at which the accumulated number of non-repeating sources becomes comparable to the total number of the repeating sources, which is a good proxy to denote the intrinsic repeater fraction among FRBs. Assuming that both types of sources exist and that their burst energies follow power law distributions, we investigate how the {\em observed} repeater fraction evolves with time for different parameters. If the lifetime of repeaters is sufficiently long so that the evolutionary effect can be neglected within the observational time span, unless $T_c \rightarrow \infty$ (i.e. there is no genuine non-repeating FRB source) the observed repeater fraction should increase with time first, reach a peak, and then decline. The peak time $T_p$ and the peak fraction $F_{\rm r,obs,p}$ depend on $T_c$ and other repeating rate parameters. With the current data, we pose a lower limit $T_c > 0.1$ d  for reasonable parameter values. We predict that future continuous monitoring of FRBs with CHIME or similar wide-field radio telescopes would obtain an $F_{\rm r,obs}$ less than $0.04$. The detection of a smaller peak value $F_{\rm r,obs,p}<0.04$ in the near future would disfavor the ansatz that ``all FRB sources repeat''. 
\end{abstract}


\section{introduction} 
Fast radio bursts (FRBs) are mysterious transients originating from distant universe \citep{lorimer07,thornton13,petroff19,cordes19}. Even though early observations only detected one-off events, the discovery of multiple bursts from FRB 121102 \citep{spitler16} suggested that at least some are repeating sources. Recent observations by CHIME revealed that repeating FRBs are commonly observed \citep{chime19a,chime19b}. 

One intriguing question is whether there exist genuinely non-repeating FRBs. The difficulty in addressing this problems lies in the wide range of the repeating waiting times \citep{palaniswamy18,caleb19} and FRB luminosities  \citep{luo18,luo20,lu19}. It is highly likely that some apparently non-repeating FRBs are actually repeaters. The non-detection of the repeated bursts could be because of the long waiting time or low flux of the repeating bursts \citep{palaniswamy18}. The high event rate of FRBs may suggest that the majority of FRBs are repeaters \citep{ravi19}. On the other hand, deep follow-up observations of some FRBs (e.g. the famous ``Lorimer'' event) did not reveal repeated bursts \citep{petroff15}. \cite{palaniswamy18} and \cite{caleb19} argued that active repeaters such as FRB 121102 are abnormally active. If all FRBs are similar to FRB 121102, many one-off bursts should have been detected as repeaters. 

However, repeating FRBs may have different degrees of repetition level. If FRBs have a wide range of repetition rate, it is possible that the anzats ``all FRB sources repeat'' is true \citep[e.g.][]{lu20}. Are there indeed genuine non-repeaters? If so, how can we find out that they exist? What are the true fractions of repeaters and non-repeaters?

In this {\em Letter}, we attempt to address these questions through Monte Carlo simulations. The basic formalism of our approach is described in Section \ref{sec:formalism}. The simulation methodology is outlined in Section \ref{sec:simulations} and the results are presented in Section \ref{sec:results}. Section \ref{sec:summary} presents conclusions with some discussion.

\section{Basic Formalism}\label{sec:formalism}
A repeating FRB source can produce a sequence of bursts, but only those with energy exceeding a threshold value are detectable. Below, we follow \cite{lu20} to calculate the repeating rate of repeaters but add a population of genuine non-repeaters. We assume that the threshold fluence for a detector to trigger an FRB event is $F_{\rm th}$. For a source at the luminosity distance $D_L$, the threshold energy for a burst to be detectable is
\begin{eqnarray}
E_{\rm th}=4\pi D_L^2F_{\rm th}(1+z)^{\alpha-1},
\label{eq:Eth}
\end{eqnarray}
where an $k$-correction has been introduced and $\alpha$ is the intrinsic spectral index of the burst ($S_\nu \propto \nu^{-\alpha}$), with $\alpha=1.5$ adopted. 

Consider that the repeating rate of a repeating FRB source is related to the intrinsic energies of the bursts as a power-law function, i.e.
\begin{eqnarray}
{dr \over dE}={r_0 \over E_0}\left({E \over E_0}\right)^{-\gamma}{\rm exp}\left(-{E \over E_{\rm max}}\right)
\label{eq:drate}
\end{eqnarray}
where $\gamma=1.92$ is inferred from the current repeating FRB sample \citep{lu20}. We set $E_0=10^{30} \ {\rm erg ~Hz^{-1}}$ and $E_{\rm max}=10^{34} \ {\rm erg~Hz^{-1}}$ \citep{lu19}. Here $r_0$ is a normalization parameter, which stands for the intrinsic repeating rate of the bursts at $E=E_0$. For FRB121102, the measured value is $r_0=0.1{\rm hr^{-1}}$  \citep{law17,james19}. Then, the effective repeating rate of bursts above $E_{\rm th}$ could be calculated as
\begin{eqnarray}
r(>E_{th})=r_0 \int_{x_{th}}^{\infty} dx x^{-\gamma} {\rm exp}\left({-x E_0 \over E_{\rm max}}\right).
\label{eq:rate}
\end{eqnarray}
where $x=E/E_0$ and $x_{th}=E_{th}/E_0$. In view of the detection of the Galactic FRB 200428 \citep{CHIME20,Bochenek20}, we also introduce a low-energy cutoff energy at $E_{\rm min}=10^{28} \ {\rm erg~Hz^{-1}}$.

As a function of $r$, the distribution of time intervals ($\delta$) between two adjacent bursts could be described by a Weibull probability density function \citep{oppermann18}, which reads
\begin{eqnarray}
{\cal W}(\delta | k,r)=k \delta^{-1}\left[\delta r \Gamma(1+1/k)\right]^k e^{-\left[\delta r \Gamma(1+1/k)\right]^k},
\label{eq:Weibull}
\end{eqnarray}
where $k$ is the shape parameter. When $k=1$, the time interval distribution reduces to the exponential distribution; when $k<1$, the bursts are clustered; and when $k>1$, the bursts tend to be periodic. A detailed discussion about the time interval distribution is shown in the Appendix.  

For non-repeating FRBs, to be detectable, they should also exceed the threshold energy shown in Equation \ref{eq:Eth}. Assume that the energy of non-repeating FRBs follow a simple power-low distribution as
\begin{eqnarray}
{dN \over dE}\propto E^{-\gamma_n}
\label{eq:dNdE}
\end{eqnarray}
with $E_{\rm n,min}<E<E_{\rm n,max}$ and We take $E_{\rm n,min}=10^{30} \ {\rm erg~Hz^{-1}}$, $E_{\rm n,max}=10^{34} \ {\rm erg~Hz^{-1}}$ and $\gamma_n=1.8$ \citep{luo20,lu19}\footnote{We take a higher minimum energy for non-repeaters than repeaters. This is because non-repeaters are supposed to originate from catastrophic events, which likely have higher energies than repeaters in general. For repeaters, the energy of some bursts could be in principle below $E_{\rm min}$. For the observational configurations we simulate (similar to that of CHIME), these low-energy bursts are not detectable at cosmological distances. }.

\subsection{Non-evolving repeaters}

In a steady state, the birth rate and the death rate of the repeating sources would balance each other, so that the total number of sources in the sky would be a constant.
Assuming that the lifetimes of the repeaters are much longer than the observational timescale and that the bursts from each source are produced with a constant repeating rate, one can approximately regard that the repeaters are not evolving. The intrinsic repeating rate for each individual repeater may not be the same, but follow a certain distribution. In this subsection, we deal with this case and defer the evolving case to the next subsection.

For genuine non-repeating sources, the progenitor of the FRB produces an FRB once in its lifetime. The number of non-repeating sources accumulate linearly with time with a constant event rate density. Let us denote the total number of repeating sources in the universe as $N_r$ and the total event rate of non-repeating FRBs in the universe  as $\dot N_n$. One can define a characteristic timescale
\begin{equation}
    T_c \equiv \frac{N_r}{\dot N_n},
\label{eq:Tc}
\end{equation}
at which the number of repeating and non-repeating sources in the sky become comparable\footnote{One may also define $T_c \equiv \rho_r / \dot \rho_n$, where $\rho_r$ is the local density of repeating sources and $\dot \rho_n$ as the local event rate density of non-repeating sources. The following discussion remains the same if the redshift distributions of the repeating sources and the non-repeating sources are roughly the same.}. 

Depending on the effective repeating rate $r$, a repeating source may be recognized as a repeater (if $r$ is large enough), an apparent non-repeater (if $r$ is smaller), or not detected at all (if $r$ is extremely small). We use $f_{\rm rr}$ and $f_{\rm rn}$ to denote the fractions of repeating sources being recognized as repeating and non-repeating sources, respectively. For non-repeating sources, only a fraction, $f_{\rm nn}$, are detected with a fluence above $F_{th}$. Therefore, for a sample of observed FRB sources, the fraction of identified repeating sources among all the detected sources should be
\begin{eqnarray}
F_{\rm r,obs}(t)&=&{f_{\rm rr}(t) N_r {\Omega \over 4\pi} \over f_{\rm rr}(t) N_r {\Omega \over 4\pi}+f_{\rm rn}(t)N_r {\Omega \over 4\pi}+f_{\rm nn}\dot{N_n}t{\Omega' \over 4\pi}} \nonumber \\
&=&{f_{\rm rr}(t) \over f_{\rm rr}(t)+ f_{\rm rn}(t)+f_{\rm nn} {t \over T_c}{\Omega' \over \Omega}},
\label{eq:frobs}
\end{eqnarray}
which is a function of the observational time $t$. Here $\Omega'$ is the field of view of the telescope, and $\Omega$ is the total sky solid angle the telescope can cover. 

\subsection{Evolving repeaters}

The evolution of repeaters has two meanings: 1) Their intrinsic repeating rates evolve with time; 2) Old sources die and new sources are born all the time in the sky. If the characteristic timescales for these evolution effects are not much longer than the observational timescale, these evolution effects should be considered.

For the repeating rates, we assume that any individual repeater is born with an intrinsic repeating rate $r_0=r_{\rm 0,max}$, which decreases with time as the source ages and finally reaches $r_0=r_{0,min}$ as the source dies. Assume all the repeaters have the same lifetime denoted as $T_l$. In a steady state, the birth rate and the death rate of the repeating sources would balance each other, so that the total number of sources in the sky would be a constant at any time. But during the observational timescale, $N_r$ would be an accumulated number.

In principle, equation \ref{eq:frobs} is still valid under these assumptions, even though the evolution is included. Since a constant parameter is needed to describe the true fraction of repeaters while $T_c$ in this case would be a function of time, we adjust the definition of $T_c$ slightly. Here we use $T_c'$ to represent the modified characteristic timescale, which reads 
\begin{eqnarray}
T_c' = {N_{r,0} \over \dot{N_n}}=T_c \times {1\over F},
\label{eq:Tcp}
\end{eqnarray}
where $N_{r,0}$ represents the total number of repeating sources at a specific time and $F = {N_{r} / N_{r,0} }$ is a  factor for evolution. Consequently, we should also replace $f_{\rm rr}$ and $f_{\rm rn}$ with $f_{\rm rr}'=f_{\rm rr} F$ and $f_{\rm rn}'=f_{\rm rn} F$. Hence we rewrite equation \ref{eq:frobs} as 
\begin{eqnarray}
F_{\rm r,obs}(t) = {f_{\rm rr}'(t) \over f_{\rm rr}'(t)+ f_{\rm rn}'(t)+f_{\rm nn} {t \over T_c'}{\Omega' \over \Omega}}.
\label{eq:frobs'}
\end{eqnarray}
When the lifetime of sources is much longer than the observational timescale ($T_l \gg t$), it would reduce to the non-evolving approximation with $F\sim 1$. Hereafter, we only distinguish $T_c$ and $T_c'$ when necessary, but only use $T_c$ when discussing them together. So do the notations $f_{\rm rr}$, $f_{\rm rn}$, $f'_{\rm rr}$ and $f'_{\rm rn}$.

\section{Monte Carlo simulations}\label{sec:simulations}

In this section, we use Monte Carlo simulations to estimate $f_{\rm rr}$, $f_{\rm rn}$ and $f_{\rm nn}$ and then predict the observed fraction of repeating bursts $F_{\rm r,obs}(t)$ at a certain observational time $t$. The results depend on several  parameters, such as $F_{\rm th}$, $T_l$, $T_c$ and $r_0$. We consider the repeating sources under both the ``non-evolving'' and ``evolving'' assumptions separately.

\subsection{Non-evolving repeaters}\label{sec:MCconst}
For non-evolving repeaters, we generate the time sequences of bursts following four steps:

\begin{itemize}
    \item Generate a series of repeating sources with a certain redshift distribution.
    \item Assign each repeating source an intrinsic repeating rate $r_0$.
    \item Calculate the effective repeating rate for each simulated repeating source.
    \item Generate burst time intervals according to the Weibull distribution (given a particular $k$ value) and form a time sequence of the repeated bursts from each repeating FRB source.

\end{itemize}

In our simulations, we assume that the redshift distributions for both repeating and non-repeating sources follow the star formation rate (SFR) history. We adopt an analytical fitting formula given by \cite{yuksel08}
\begin{eqnarray}
{\rm SFR}(z)\propto\left [(1+z)^{3.4\eta}+\left (\frac{1+z}{5000}\right )^{-0.3\eta}+\left (\frac{1+z}{9}\right )^{-3.5\eta}\right ]^{1/\eta},
\label{eq:SFR}
\end{eqnarray}
where $\eta=-10$. We generate $N_r$ redshift values according to Equation \ref{eq:SFR} and assign each value to one repeating FRB source. Choosing appropriate values of $F_{th}$ and $r_0$, for each repeater, we calculate its effective repeating rate $r$ using Equation \ref{eq:Eth} - \ref{eq:rate}. 

Assume the observation starts from $T_0$. A waiting time needs to be introduced to describe how long it  takes to detect the first burst from each source. In our simulations, we randomly generate a time interval $\delta_1$ according to Equation \ref{eq:Weibull} and then randomly generate a waiting time $t_w$ in the range of [0,~$\delta_1$]\footnote{When $k=1$ in Equation \ref{eq:Weibull} is assumed, the distribution of $t_w$ is the same as that of $\delta$.}. Hence, the first burst appears at $T_1=T_0+t_w$. For the sake of convenience, we set $T_0=0$. For each repeating source, we simulate $N$ bursts, which appear at $T_i=T_1+\sum_{j=2}^{i} \delta_j$ where $i=1,2,3...N.$. 

\subsection{Evolving repeaters}

The main procedure to generate the time sequences of bursts for evolving repeaters follows five steps:

\begin{itemize}
    \item Generate a series of repeating sources with a certain redshift distribution.
    \item Assign each source an intrinsic repeating rate ($r_0$) and an age ($T_a$).  
    \item Calculate the effective repeating rate for each simulated repeating source.
    \item Generate one burst time interval for each repeater according to the Weibull distribution (given a particular $k$ value). Update the intrinsic repeating rate and age. 
    \item Repeat the previous step until the end of time sequence exceeds the lifetime of repeater. Then replace this source with a newly born one.
\end{itemize}

Under the assumption that the total number of repeaters in the sky at an arbitrary specific time is unchanged because the birth and death rates balance each other, the age of repeaters ($T_a$) in the sky should distribute uniformly from $0$ to their lifetimes. Generate a series of $T_a$ randomly and assign to each source. For each source, every time when a burst is produced, we check if $T_i$ exceeds the lifetime $T_l$. If so, we record and stop this time sequence and replace it with a new one. Set the new one with the same redshift as the dead one, with $T_a=0$. The time sequence of the new source's bursts starts from ($T_l-T_{a,i}$).
The evolution of $r_0$ could be introduced when the function of $dr_0 / dt$ is given.
Update $r_0$ every time a new time interval is produced. 

In principle, we can evolve $r_0$ and $T_a$ independently. With a similar method as that in section \ref{sec:MCconst}, we can obtain a time sequence $T_i$ for each repeater under the evolving assumption.

\subsection{Observational configuration}

Since the bursts are simulated with an effective repeating rate $r$ (Equation \ref{eq:rate}), all of them are above the flux threshold of a telescope. Those fall into the field of view of the telescope at the burst time would be detected.
In reality, the telescope may not stare at one particular sky area all the time. Therefore, for each repeating source, the observation is not continuous, but consists of a number of discrete short-term observations. We assume $t_e$ as the duration of each observation at a certain sky area, and $t_g$ as the gap between two observations at the same area. For an observing time $t$ ($t<T_{i=N}$) for a telescope, there are $n=t/(t_e+t_g)$ observing periods. From a simulation, we count the number of $T_i$'s for each source that satisfy $m(t_e+t_g)<T_i<t_e+m(t_e+t_g)$,
where $m=0,1,2...n-1$. If more than one burst was detected, it would be recognized as a repeater; if exactly one burst was detected, it would be recognized as an apparently non-repeating source; otherwise, the source would not be detected. Investigating all the sources in the simulation, we can obtain the values of $f_{\rm rr}$ and $f_{\rm rn}$. 

Non-repeating sources are generated following the redshift (Equation \ref{eq:SFR}) and energy (Equation \ref{eq:dNdE}) distributions. The fraction $f_{\rm nn}$ could be estimated through dividing the number of detectable non-repeating bursts by the total number of simulated non-repeating sources. Note that $f_{\rm nn}$ is not a function of time.

In our simulations, we take the threshold fluence as $F_{\rm th}=4 \ {\rm Jy \ ms}$, which is comparable to CHIME's sensitivity \citep{chime19b,fonseca20}. We generate $10^6$ non-repeating sources with $E_{\rm n,min}<E<E_{\rm n,man}$, so that $f_{\rm nn}$ can be determined as $\sim 0.0035$. According to our definition, the total solid angle $\Omega$ covered by the telescope would be observed once in $t_e+t_g$, and $t_e$ is the effective observational time for the field of view with a solid angle $\Omega'$. We then have $\Omega'/\Omega = n t_e/t$. Equations \ref{eq:frobs} and \ref{eq:frobs'} can be then rewritten as
\begin{eqnarray}
F_{\rm r,obs}={f_{\rm rr}(t) \over f_{\rm rr}(t)+ f_{\rm rn}(t)+f_{\rm nn} {nt_e \over T_c}}
\end{eqnarray}
and
\begin{eqnarray}
F_{\rm r,obs}={f_{\rm rr}'(t) \over f_{\rm rr}'(t)+ f_{\rm rn}'(t)+f_{\rm nn} {nt_e \over T_c}}
\end{eqnarray}

\section{Results}\label{sec:results}
\subsection{Evolution of $F_{\rm r,obs}$}\label{sec:evolution}

Our first goal is to investigate how the observed repeater fraction, $F_{\rm r,obs}$, evolves with time, and how this evolution depends on the parameter $T_c$, a characteristic parameter to define the relative fractions between the genuine repeaters and non-repeaters: $T_c~ \rightarrow \infty$ means all FRBs are  repeaters, and $T_c~ \rightarrow 0$ means that the majority FRBs are genuine non-repeaters. 
\subsubsection{Non-evolving repeaters}

We first give an example by assuming that all repeaters are as active as the first repeating source FRB 121102 ($r_0=0.1 {\rm hr^{-1}}$ \citep{law17,james19} and last forever. The observed fraction of repeating sources $F_{\rm r,obs}$ as a function of observational time $t$ is shown in the upper panel of Figure \ref{fig:r0m1}. For $T_c\rightarrow \infty$, $F_{\rm r,obs}$ always increases, but the slope decreases as a function of time. If $T_c$ is a finite value, which means that there are genuine non-repeating sources, $F_{\rm r,obs}$ increases with time first because of the fast increasing of $f_{\rm rr}$ in the beginning.
Later on, the increase of $f_{\rm rr}$ slows down because one already recognizes most of the repeaters. On the other hand, the number of non-repeating sources linearly increases with time, so that $F_{\rm r,obs}$ would reach a peak and then starts to decline afterwards. 

In the following, we denote the expected maximum fraction of repeating sources as the ``peak fraction" ($F_{\rm r,obs,p}$) and its corresponding time is expressed as the ``peak time" ($T_p$).

From the upper panel of Figure \ref{fig:r0m1}, we can see that the distinction of $F_{\rm r,obs}$ curves for different $T_c$ is insignificant when $t$ is short. A distinct feature occurs around the peak time. The peak time and peak fraction are the most crucial observational quantities that can be used to estimate $T_c$. A smaller $T_c$ corresponds to a more dominant non-repeater population, which corresponds to a smaller peak time and a lower peak fraction. 

\subsubsection{Evolving repeaters}
\begin{figure}[ht!]
\resizebox{90mm}{!}{\includegraphics[]{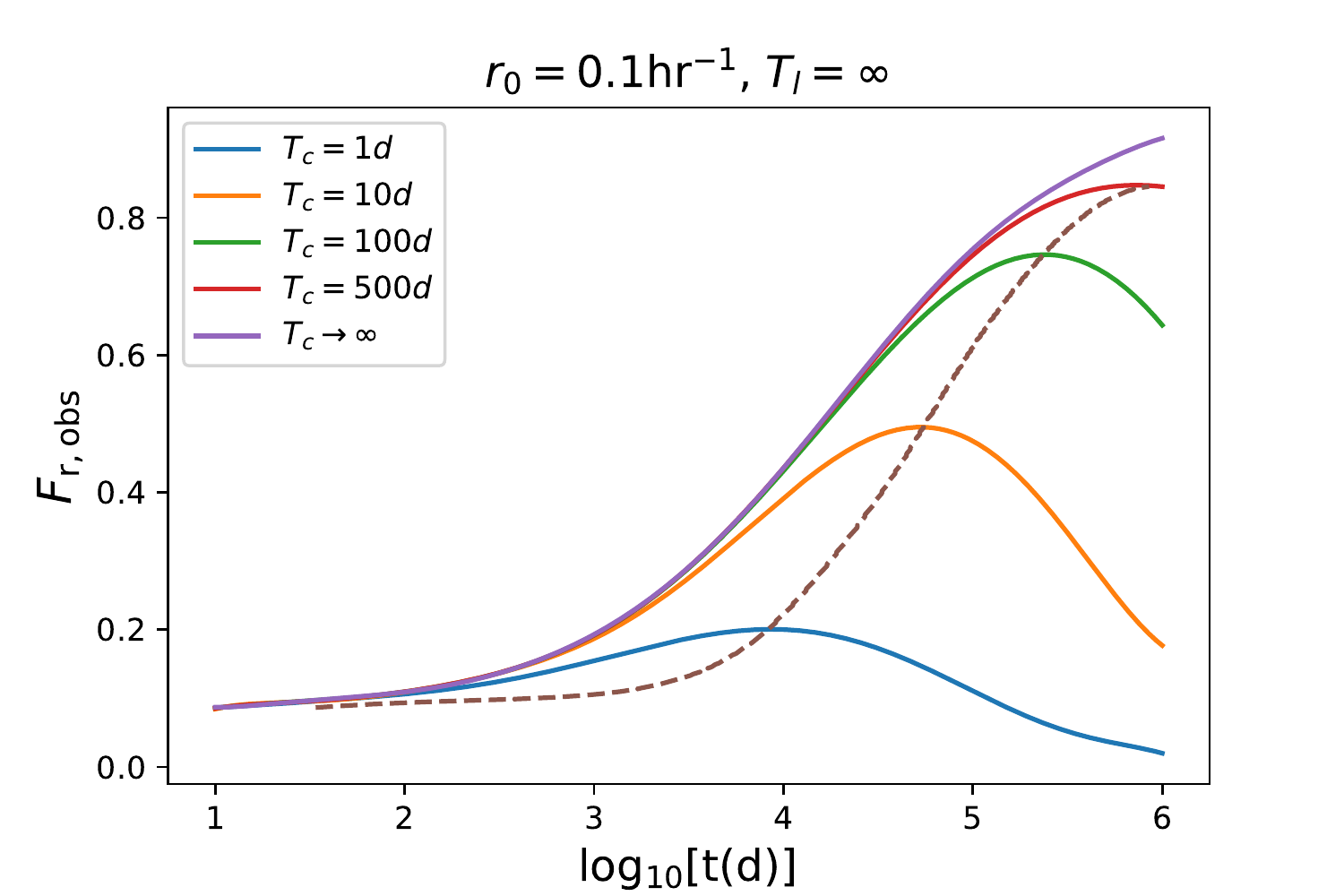}}
\resizebox{90mm}{!}{\includegraphics[]{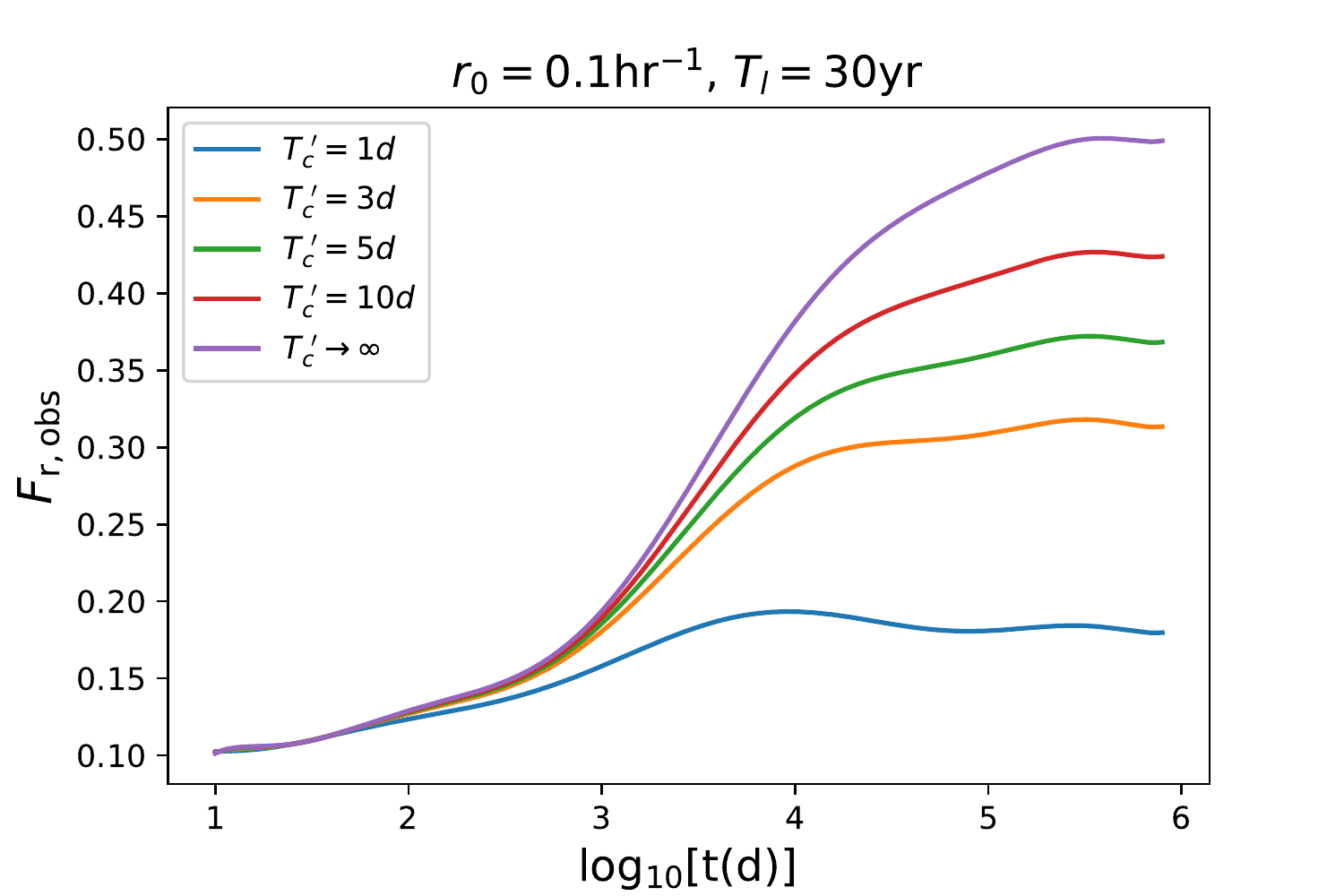}}
\caption{The observed repeater fraction $F_{\rm r,obs}$ as a function of time. The colored solid lines delineate the evolution of $F_{\rm r,obs}$ for different $T_c$ values. Upper panel: Constant repeating rate ($r_0=0.1~hr^{-1}$) with infinity lifetime. The dashed line marks the trajectory of the peak fraction vs. peak time as a function of $T_c$.
Lower panel: Constant repeating rate ($r_0=0.1~hr^{-1}$) with lifetime $T_l=30 {\rm yr}$. Following parameters are adopted for both panels : $t=n(t_e+t_g)$, where $t_e=0.2~{\rm hr}$ and $t_g=23.8~{\rm hr}$; the Weibull parameter is adopted as $k=0.3$ \citep{oppermann18}.}
\label{fig:r0m1}
\end{figure}

Keep $r_0=0.1 {\rm hr^{-1}}$ and set a finite lifetime as $T_l=30{\rm yr}$ to all the repeaters. The evolution of $F_{\rm r,obs}$ is shown in the lower panel. The curves flatten when $t \gg 30$ yr. Because the observational timescale can totally cover the entire lifetime of the repeaters, both repeating and non-repeating bursts would accumulate linearly with time with constant rates. Therefore, $F_{\rm r,obs}$ may never approach $1$, but balance at a certain level smaller than $1$, even if $T'_c \rightarrow \infty$. If $T'_c$ is small, a peak of $F_{\rm r,obs}$ could be observed at $T_p<T_l$. Hence, with a large $T_l$, a peak would still be expected. See the lower panel of Figure \ref{fig:r0m1}.

\subsection{Effect of key parameters}

As already known from section \ref{sec:evolution}, with a fixed $r_0$, $T_c$ is a parameter that  strongly influences $T_p$ and $F_{\rm r,obs,p}$. In this section, we allow the value of $r_0$ vary among repeating FRBs to get more realistic results. 

\subsubsection{Non-evolving repeaters} \label{sec:resultcons}
Our simulation starts from non-evolving repeaters. The evolution of $r_0$ as well as the birth and death of sources are not included.
We consider that the values $r_0$ for different sources follow a power-law distribution, i.e.
\begin{equation}
    \frac{dn^*}{dr_0} \propto r_0^{-q},
\label{eq:r0}
\end{equation}
where $dn^*$ represents the number of repeating sources with the intrinsic repeating rate in the range of $r_0\sim r_0+dr_0$. For this distribution, one can introduce three parameters: two for the $r_0$ range ($r_{\rm 0,min}$ and $r_{\rm 0,max}$) and one for the power-law index ($q$). Among these three parameters, $r_{\rm 0,max}$ is the most accessible one, which could be directly measured from the most active repeating sources in the universe (e.g. active repeaters at a relatively large distance). The index $q$ can also be obtained observationally by fitting the repeaters' repetition rate distribution. However, the value of $r_{\rm min,0}$ is much more difficult to determine, because observationally it is hard to distinguish a repeating source with a very low $r_0$ from an intrinsically non-repeating burst.

In our simulation, we adopt a fixed maximum repeating rate ${\rm r_{0,max}=10^{0.5}hr^{-1}}$, which is obtained by fitting CHIME's latest repeating FRB sample \citep{lu20}. We then choose different $r_{\rm 0,min}$ and $q$ values to see how they influence the simulation results. We fix one of these two parameters and vary the other one in every simulation. The results are shown in Figure \ref{fig:FrTp} with solid lines denote the locus of $(T_p, F_{\rm r,obs,p})$ as the parameters are varied. When we set a fixed index as $q=1.6$ and vary the value of $r_{\rm 0,min}$, the results are shown in the left panel of Figure \ref{fig:FrTp}. With the same $T_c$, a lower $r_{\rm 0,min}$ would lead to a smaller $F_{\rm r,obs,p}$ and a smaller $T_p$. When we set $r_{\rm 0,min}=10^{-5.5}{\rm hr^{-1}}$ and vary the value of $q$, the results are shown in the right panel of Figure \ref{fig:FrTp}. With the same $T_c$, a higher $q$ would lead to a smaller $F_{\rm r,obs,p}$ and a smaller $T_p$. 

Both a lower $r_{\rm 0,min}$ and a higher $q$ would make more repeating sources with low $r_0$ values. Consider that the increase of $f_{\rm rr}$ slows down when most of the sources which repeat frequently enough have been recognized as repeating sources. If there are more low $r_0$ sources, $f_{\rm rr}$ would have a smaller absolute value and its increase rate would become smaller. This explains why both $F_{\rm r,obs,p}$ and $T_p$ become smaller in these cases.

\subsubsection{Evolving repeaters}
\label{sec:resultsev}
In this section, we set $r_0$ and $T_a$ to have one-to-one correspondence. As assumed above, the birth and death of repeaters balance each other, which means that their age distribution is stable, thus $r_0$ distribution should also be stable. We then assume that all the repeaters are born with $r_0=r_{\rm 0,max}$ and die with $r_0=r_{\rm 0,min}$ with the same evolutionary track, which can be derived from the $r_0$ distribution at an arbitrary time.

Using the power-law distribution in Equation \ref{eq:r0}, $T_a$ changes with $r_0$ as
\begin{eqnarray}
    \frac{dT_a}{dr_0} = { T_l(-q+1) \over  r_{0,max}^{-q+1}-r_{0,min}^{-q+1}} r_0^{-q},
\end{eqnarray}
where $T_l$ is the lifetime of the sources. 
Hence the age of a repeater could be estimated from its intrinsic repeating rate as
\begin{eqnarray}
T_a =  { r_{0,max}^{-q+1}-r_0^{-q+1} \over  r_{0,max}^{-q+1}-r_{0,min}^{-q+1}} T_l.
\end{eqnarray}
The evolution of $r_0$ follows
\begin{eqnarray}
    \frac{dr_0}{dt}= { r_{0,max}^{-q+1}-r_{0,min}^{-q+1} \over T_l(-q+1) } r_0^q.
\end{eqnarray}
Considering both the lifetime and evolution of intrinsic repeating rate, we conduct simulations with the same parameter sets used in section \ref{sec:resultcons}. We find that $T_c'$, $r_{\rm 0,min}$ and $q$ influence $F_{\rm r,obs,p}$ and $T_p$ with the same trend as that with non-evolving repeaters, which is shown in Figure \ref{fig:FrTpe}. For the same ($T_p$, $F_{\rm obs,p}$) pair, the required $T'_c$ is slightly greater than $T_c$. This is understandable because the repeating rate decreases with time under the evolving assumption, which would lead to less repeating sources to be recognized, and hence, less genuine non-repeating sources are expected. However, the difference is insignificant because the lifetime of  repeaters is assumed to be much longer than the observational timescale.

\begin{figure*}[ht!]
\begin{center}
\begin{tabular}{l}
\resizebox{90mm}{!}{\includegraphics[]{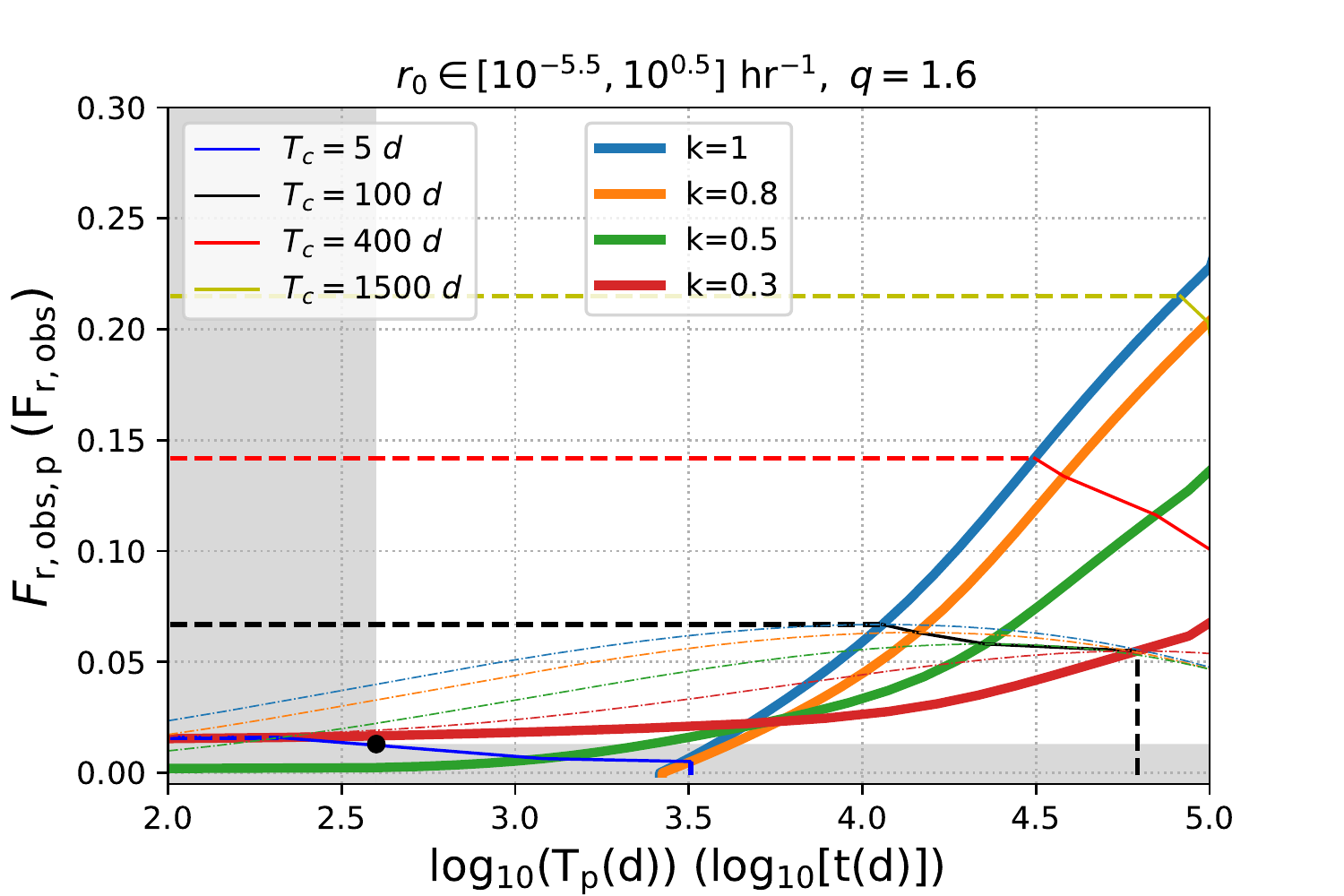}} \\
\end{tabular}
\begin{tabular}{ll}
\resizebox{90mm}{!}{\includegraphics[]{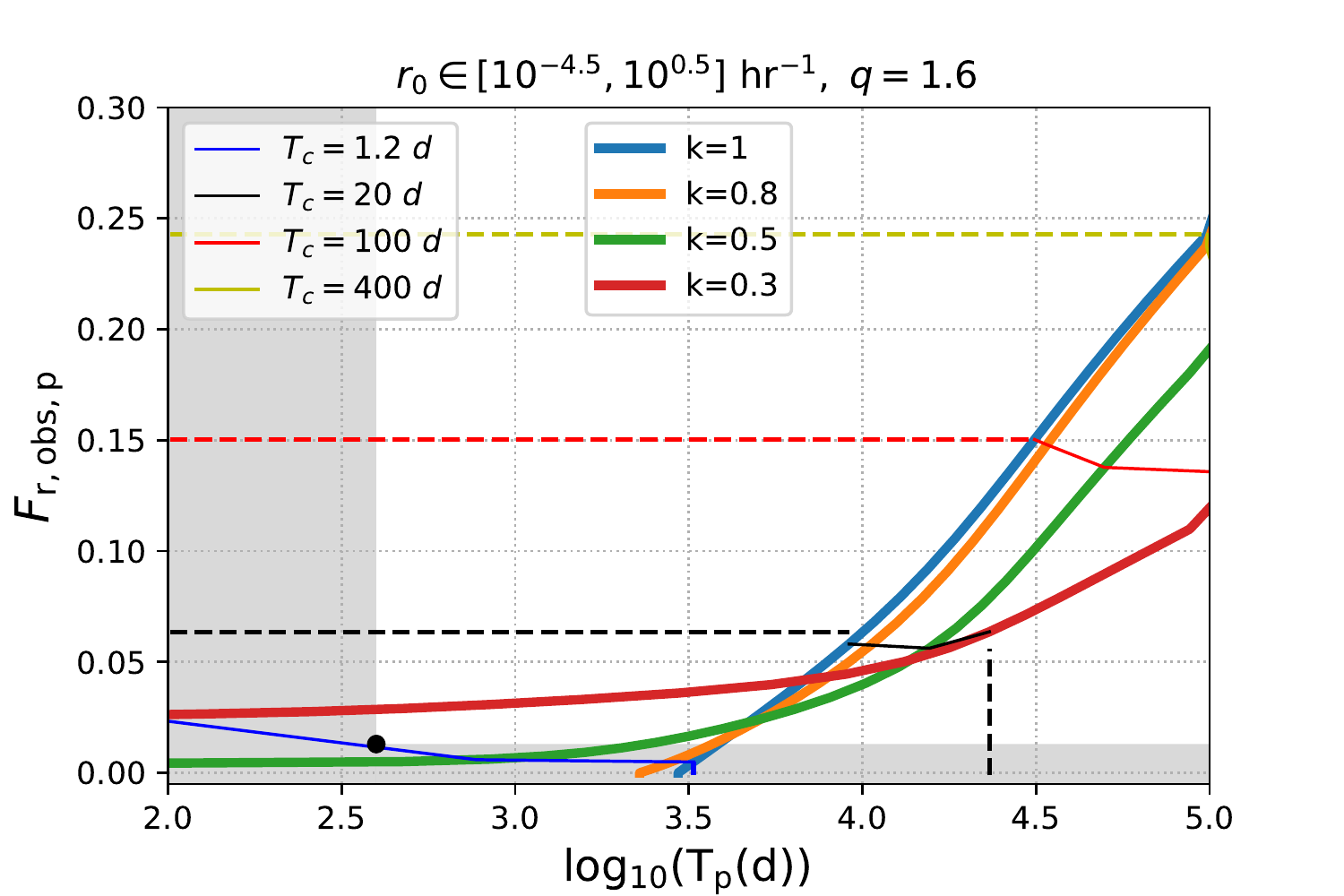}}&
\resizebox{90mm}{!}{\includegraphics[]{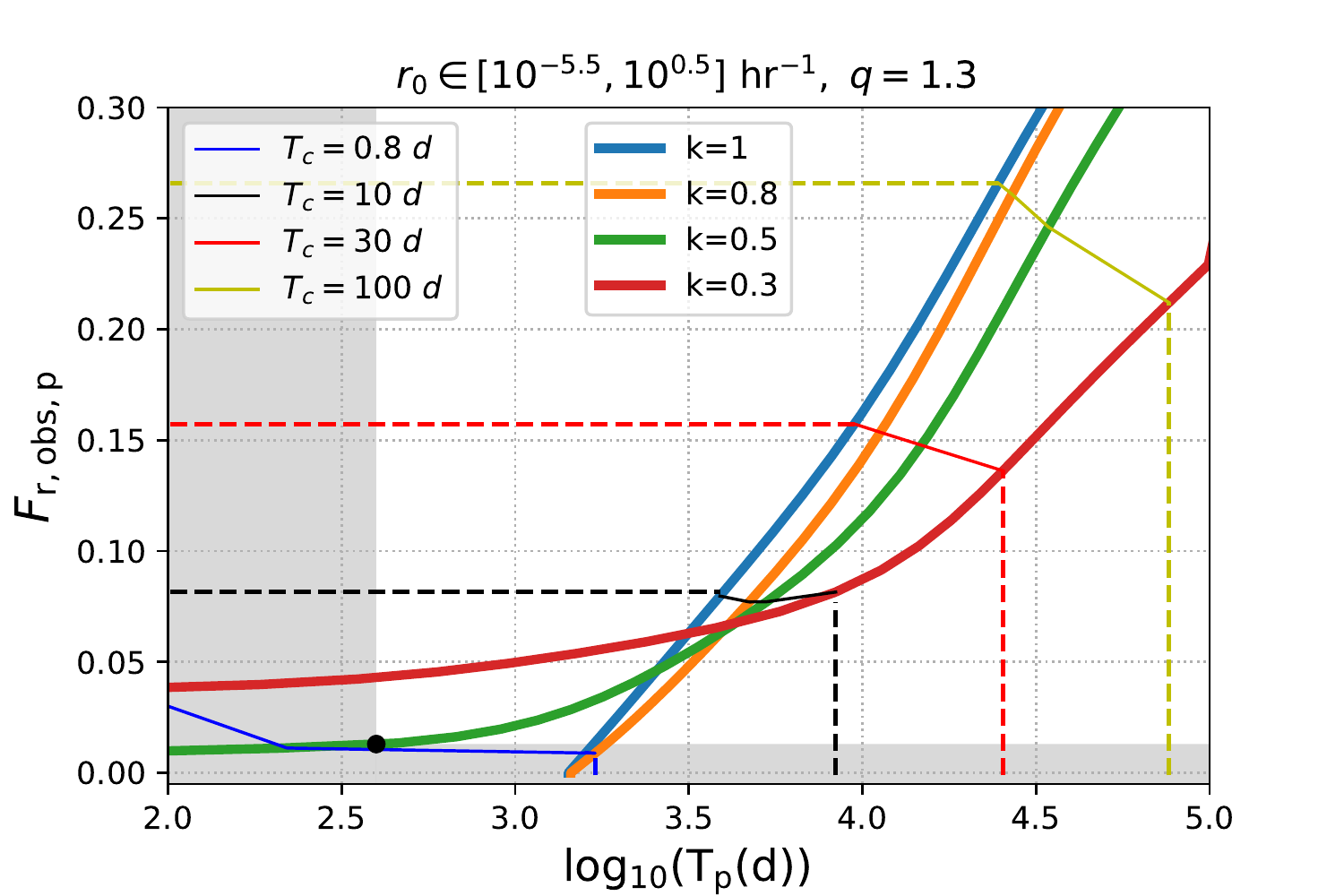}}
\\
\resizebox{90mm}{!}{\includegraphics[]{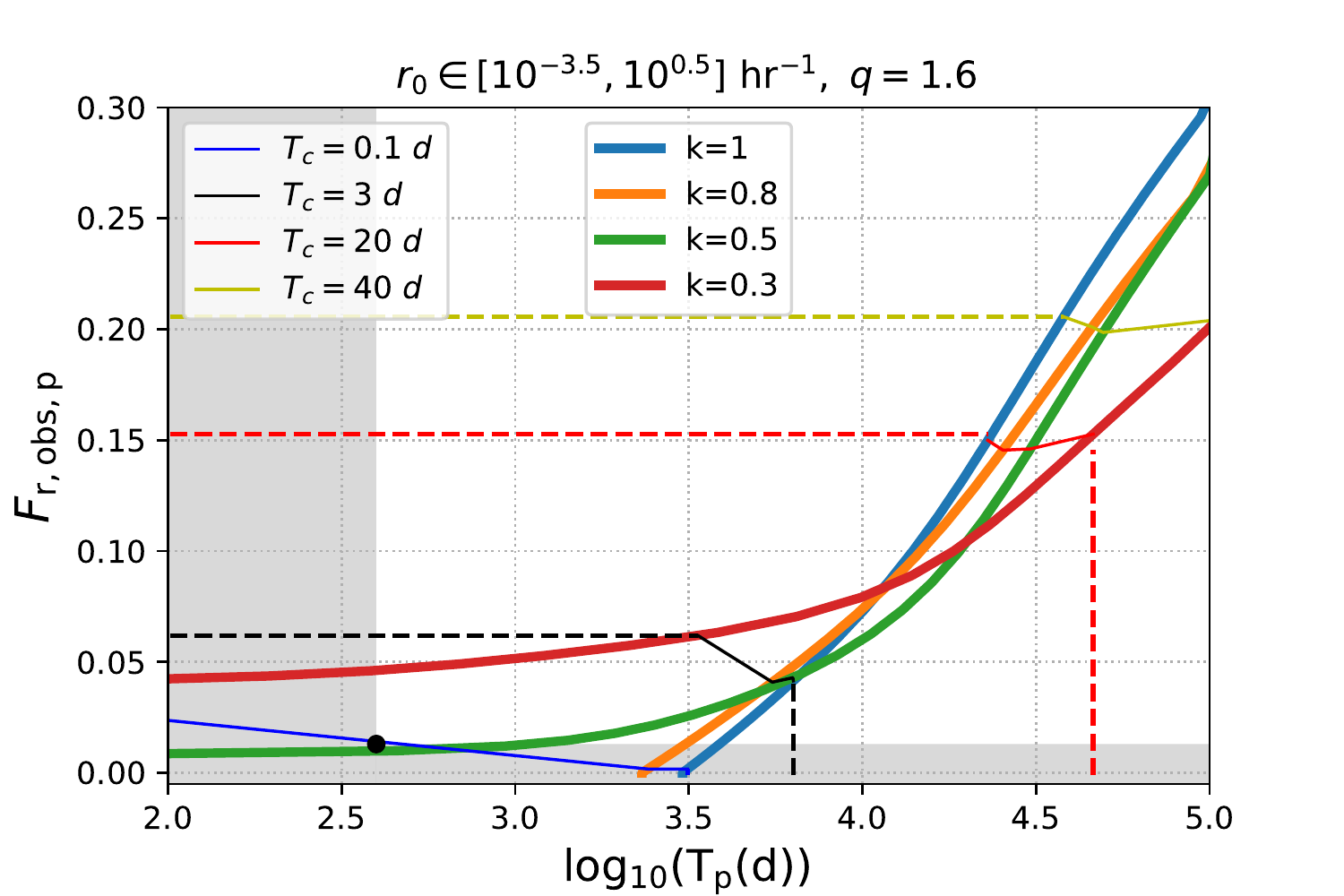}}&
\resizebox{90mm}{!}{\includegraphics[]{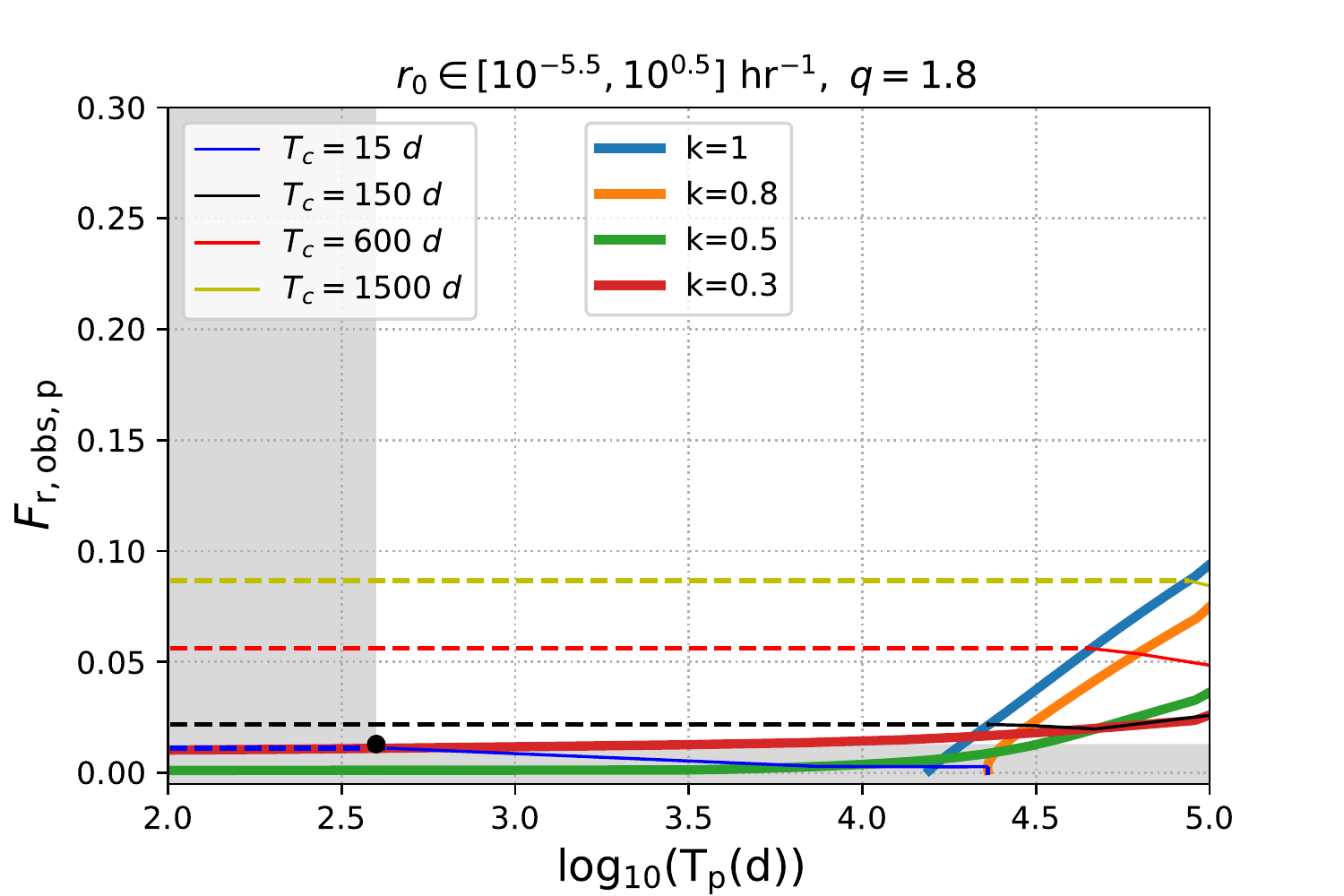}}
\end{tabular}
\caption{Observed peak repeater fraction and peak time ($T_p - F_{\rm r,obs,p}$) of repeating FRB sources for different model parameters under the ``non-evolving repeater'' assumption. All the figures allow an $r_0$ distribution among repeating FRB sources, with different $r_0$ distribution range and $q$ values marked on the top of each panel. Thick solid lines stand for the locus $(T_p, F_{\rm r,obs,p})$ when different $T_c$ (thin solid color contours) and $k$ values (thick solid color lines) are assumed. Dashed lines are the most conservative constraints on $T_c$ through the observational time $t$ and observed fraction $F_{\rm r,obs}$ before a peak is reached. In the upper panel, the dot-dashed lines show the evolution of $F_{\rm r,obs}$ as a function of time as examples. The black dot in each figure stands for the observational time and fraction of repeating sources according to the CHIME 400-d observation results \citep{fonseca20}. All five figures make the assumption of $t=n(t_e+t_g)$, where $t_e=0.2~{\rm hr}$ and $t_g=23.8~{\rm hr}$.}
\label{fig:FrTp}
\end{center}
\end{figure*}

\begin{figure*}[ht!]
\begin{center}
\begin{tabular}{l}
\resizebox{90mm}{!}{\includegraphics[]{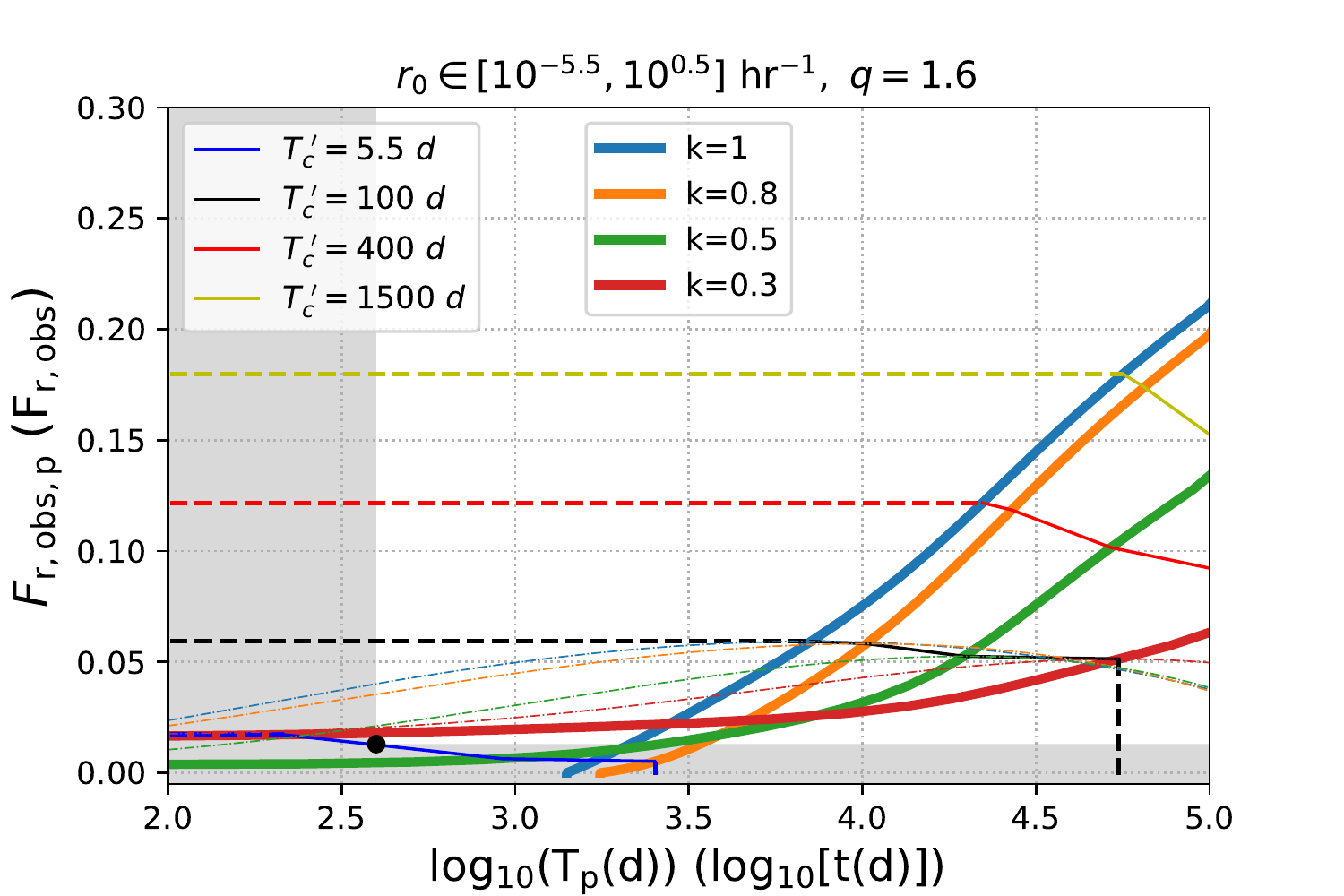}} \\
\end{tabular}
\begin{tabular}{ll}
\resizebox{90mm}{!}{\includegraphics[]{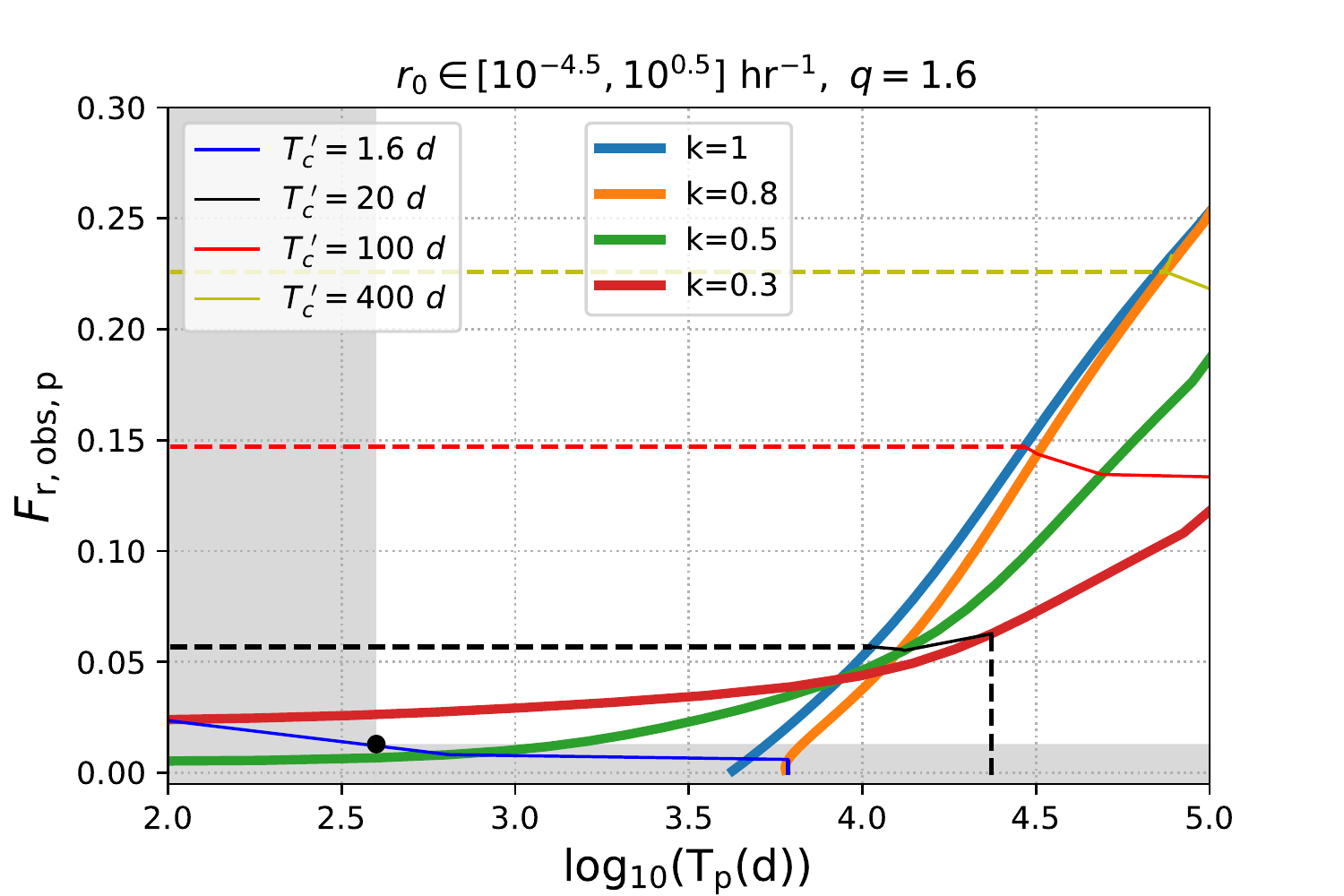}}&
\resizebox{90mm}{!}{\includegraphics[]{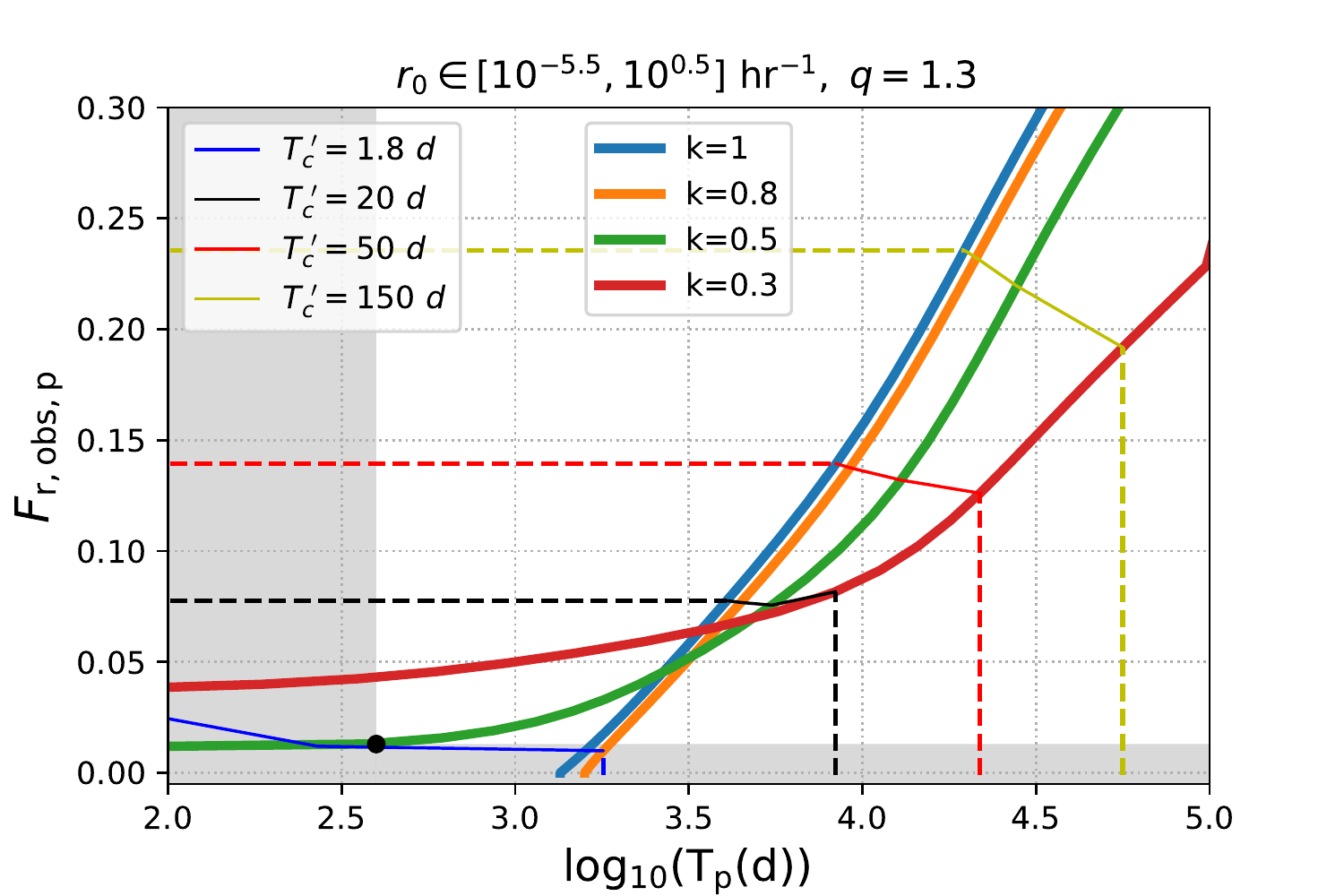}}
\\
\resizebox{90mm}{!}{\includegraphics[]{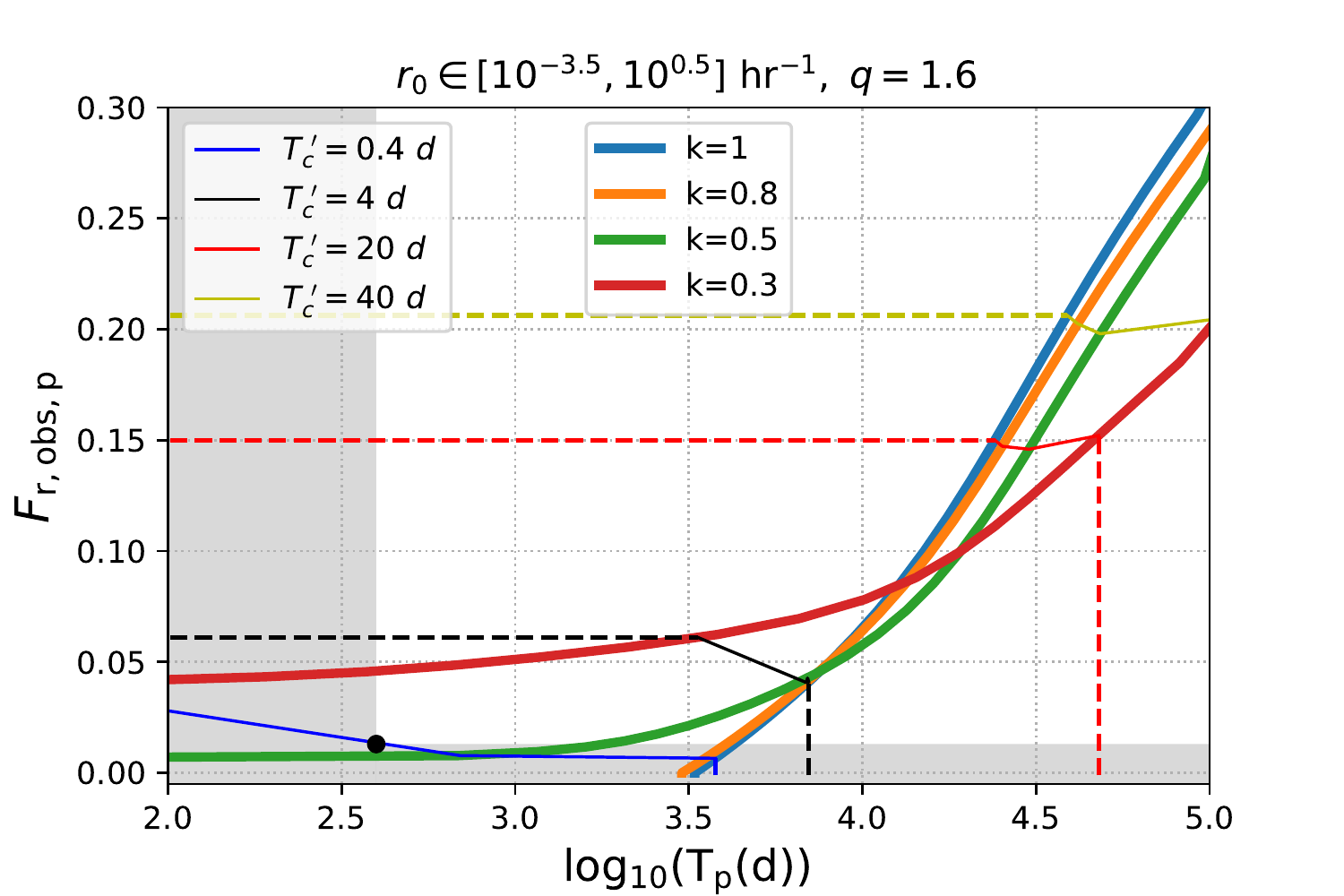}}&
\resizebox{90mm}{!}{\includegraphics[]{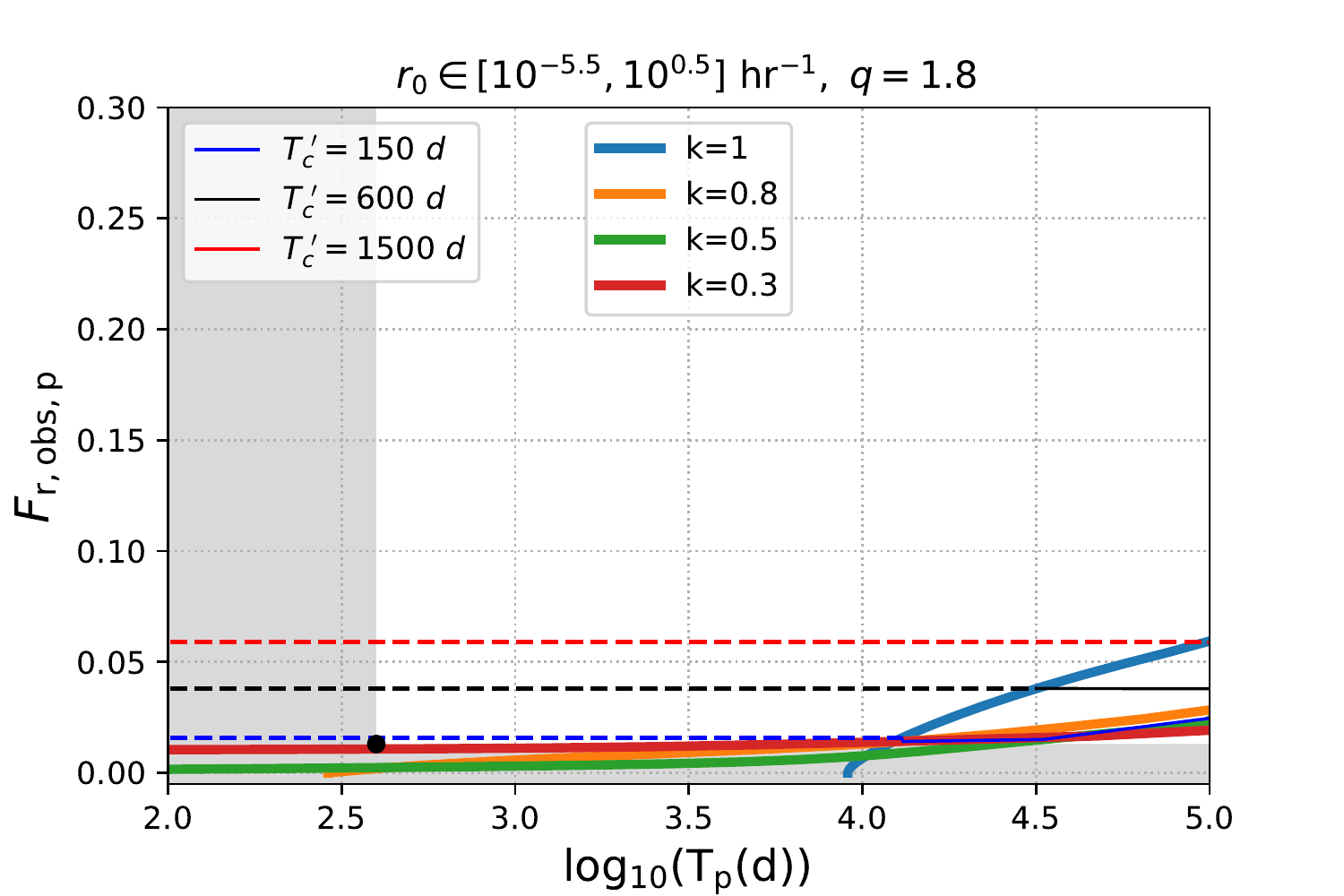}}
\end{tabular}
\caption{Observed peak repeater fraction and peak time of repeating FRB sources under the ``evolving repeater'' assumption. All the notations are the same as those of Figure \ref{fig:FrTp}.}
\label{fig:FrTpe}
\end{center}
\end{figure*}

In Figure \ref{fig:FrTp} and Figure \ref{fig:FrTpe}, we also allow that the shape parameter $k$ for the Weibull distribution to vary in different simulations. We find that the $k$ value would dramatically influence the peak time but only slightly influence the peak fraction when other parameters are set to fixed values.

\subsection{Constraints on $T_c$}
Four parameters ($T_c$, $r_{\rm 0,min}$, $q$ and $k$) have been discussed in the previous sections. The latter three parameters are related to repeating sources only, which may be eventually measured from the observations of repeaters. The $T_c$ parameter concerns the true relative fractions of repeaters and non-repeaters, which cannot be measured directly from the repeater data only. It may, however, be constrained from the observed $F_{\rm r,obs}$ as a function of time, or the measurement of ($T_p - F_{\rm r,obs,p}$) if a peak indeed exists. In principle, for a reasonable $k$ range (e.g. from $0.3$ to $1$), once a peak is detected, one may find an appropriate $r_0$ distribution to make the observed peak point located in the region of the predicted peak points within the assumed $k$ range (contours in Figure \ref{fig:FrTp} and Figure \ref{fig:FrTpe}). Finding the $T_c$ contour line that goes through the observed peak point, one can then determine both $T_c$ and $k$. In reality, there might not be only one $r_0$ distribution that can satisfy the observational constraint. Hence, constraints on $r_0$ distribution from the repeating FRB data would be helpful to make more stringent constraints on $T_c$.

According to our simulations, depending on parameters $T_p$ can be much longer than the observational timescale, e.g. up to thousands of years. In general, even when the peak (if exists), i.e. $(T_p,F_{\rm r,obs,p})$, has not been detected yet, one can still put constraints on $T_c$. Similar to the case with an observed peak, one can also find the $T_c$ value corresponding to the current $F_{\rm r,obs}$ and observational time $t$. This $T_c$ value would serve as the lower limit, since both $F_{\rm r,obs,p}$ and $T_p$ increase with $T_c$. In the case when $F_{\rm r,obs}$ and $t$ do not appear in the solid $T_c$ contour region in Figure \ref{fig:FrTp} and Figure \ref{fig:FrTpe}, we also plot the conservative constraints on $T_c$ with dashed lines in Figure \ref{fig:FrTp} and \ref{fig:FrTpe}. The observational run by CHIME from 2018 August 28 to 2019 September 30 has detected $\sim 700$ new FRBs with $9$ repeaters \citep{fonseca20}. We thus place the fraction $~0.013$ with $~400$ d observation in each panel of Figure \ref{fig:FrTp} and Figure \ref{fig:FrTpe} to denote the current data constraint.

\begin{figure}[ht!]
\resizebox{90mm}{!}{\includegraphics[]{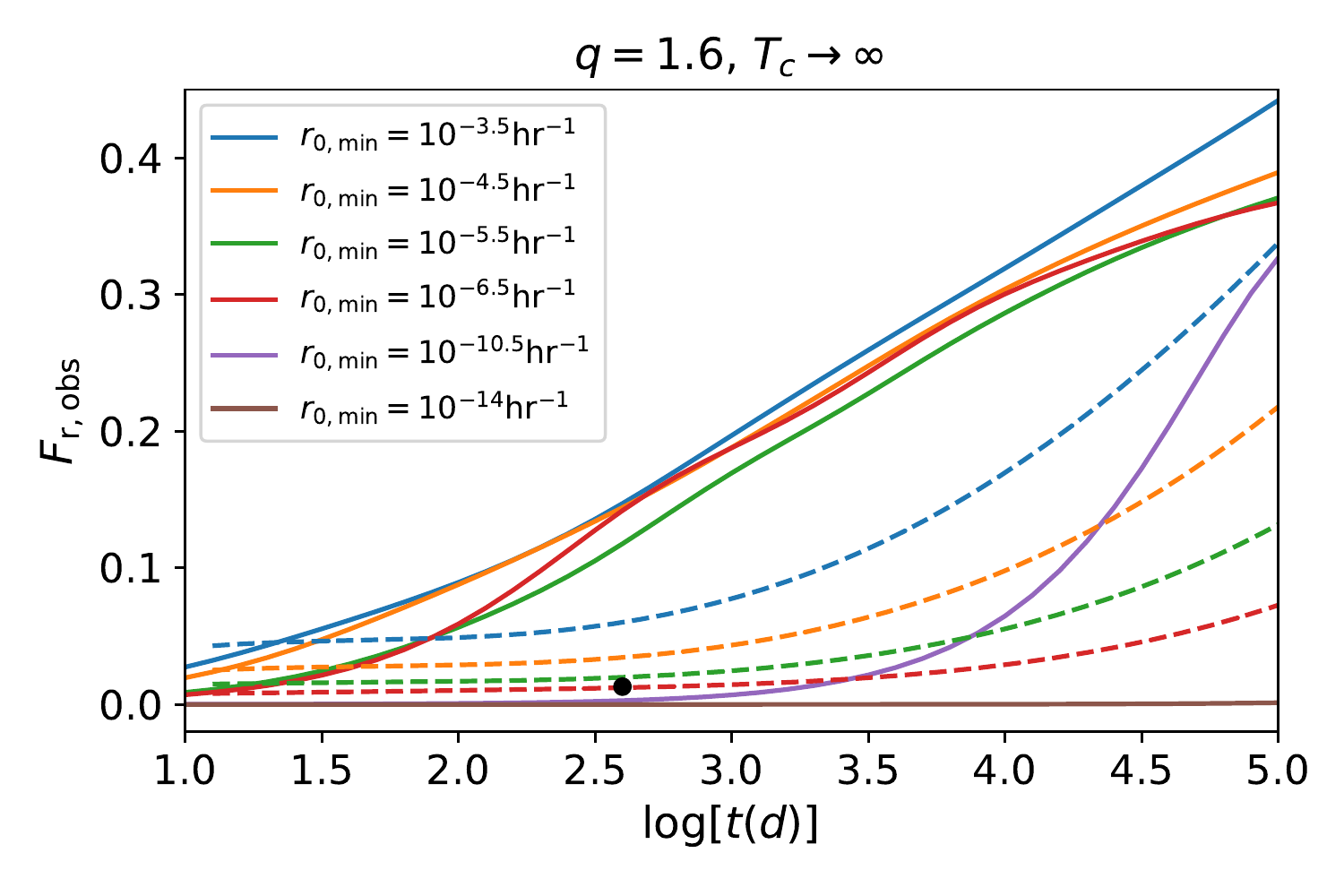}} \\
\resizebox{90mm}{!}{\includegraphics[]{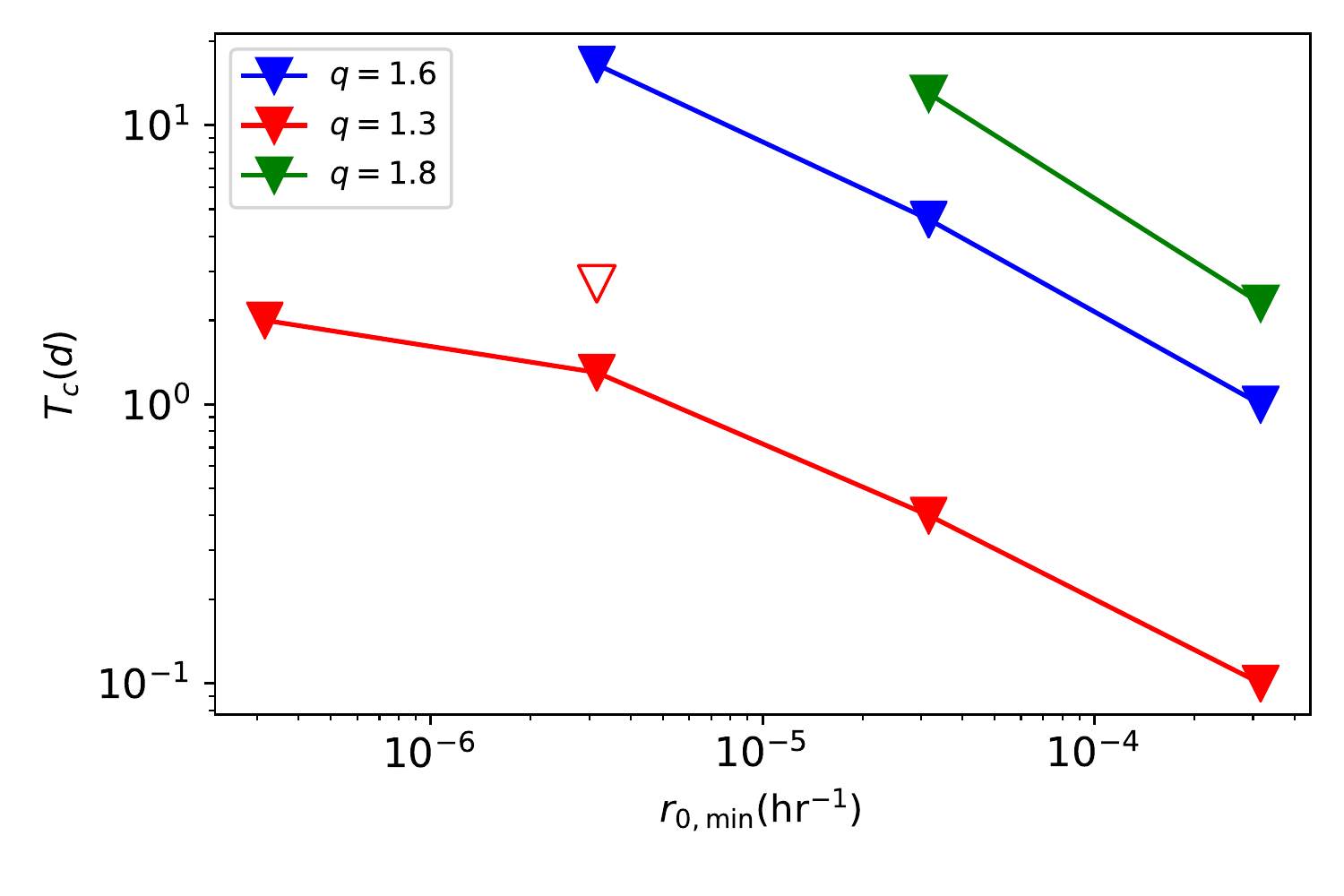}}
\caption{Upper panel: The evolution of $F_{\rm r,obs}$ with different $r_{\rm 0,min}$ values adopted. The black dot is the same as that in Figure \ref{fig:FrTp}, which represent the currents data point. The solid and dashed lines stand for the cases for $k=1$ and $k=0.3$, respectively. Lower panel: The upper limit of $T_c$ with the current observed fraction of repeaters. The minimum value of $r_0$ serves as the horizontal axis while the results with different $q$ values are shown with different colors. The solid triangles and the line stand for the non-evolving repeater case, while the hollow triangle and the dashed line stand for the evolving repeater case. The upper limits are obtained with $k=0.3-1$.} 
\label{fig:Frobs}
\end{figure}

The simulated evolution curve of $F_{\rm r,obs}$ should certainly pass through the observational value. Once the predicted $F_{\rm r,obs}$ is lower than the observed value even when $T_c \rightarrow \infty$, this parameter set of repeaters should be ruled out. If the predicted  $F_{\rm r,obs}$ is higher than the truly observed value, one can still lower the $T_c$ value to meet the data constraint. Hence, under the assumption that $r_0$ satisfies a power-law distribution, since a lower $r_{\rm 0,min}$ leads to a lower $f_{\rm rr}$ and thus a lower predicted $F_{\rm r,obs}$, the current data point could place a lower limit on $r_{\rm 0,min}$. As shown in the upper panel of Figure \ref{fig:Frobs}, for $k=0.3$ (the favored value of $k$ for FRB 121102 \citep{oppermann18}), $r_0 < 10^{-6.5}{\rm hr}$ would be disfavored, since the evolution of $F_{\rm r,obs}$ would never pass through the current data point. However, if we relax the constraints on $k$, $r_{\rm 0,min}<10^{-6.5}{\rm hr}$ would still be possible. For $k=1$, $r_{\rm 0,min}$ could be as low as $10^{-10.5}{\rm hr^{-1}}$ without violating the current data point. If $r_{\rm 0,min}=10^{-14}{\rm hr}$ is chosen, which corresponds to the Hubble timescale, the evolving curve of the observed fraction of repeating FRBs would remain near $0$ all the time, because repeating FRBs in this case is extremely hard to detect. This figure is made under the ``non-evolving'' assumption. If evolution is considered, as discussed in \ref{sec:resultsev}, the results would be very similar. Hereafter, we will not discuss "evolving" and "non-evolving" cases separately, because when the distribution of $r_0$ is considered, their results are very close. We also test other $q$ values from $1$ to $2$.\footnote{If $q>2$, most of the repeaters are located at the lower end of the $r_0$ distribution, which makes it extremely difficult to observe a burst produced from them. It is even more difficult to detect them as repeaters. This means that the observed $F_{\rm r,obs}$ would remain very small for a long time, thus it would be inefficient to pose a lower limit on $T_c$. However, a large $q$ value is already disfavored by the current date because it would not lead to a $F_{\rm r,obs}$ value as large as $0.013$. In the following, we do not consider the case with $q>2$. } It turns out that the general trend does not change with the results only slightly differ in numbers. This is reasonable because both $f_{\rm rr}$ and $f_{\rm rn}$ would decrease with a lower $r_{\rm 0,min}$ or a larger $q$. Thus, the decrease of $F_{\rm r,obs}$ would not be very significant.

On the other hand, with a certain $r_0$ distribution and a reasonable range of $k$ assumed, one can put both upper and lower limits on $T_c$. Compared with the lower limit obtained from the peak fraction constraint discussed above, the lower limit of $T_c$ here would be more conservative, because here we do not involve the assumption that the observed fraction of repeaters has not reached the peak value yet.

Considering that the current CHIME data point of the observed fraction of repeaters is relatively low, we have reasons to expected a higher $F_{\rm r,obs}$ with a longer observational timescale in the future. Even if the current data point has passed the peak fraction, the current observed fraction should be still around the peak value. Hence we adopt the lower limit obtained from Figure \ref{fig:FrTp} and Figure \ref{fig:FrTpe} while using the evolution curve of $F_{\rm r,obs}$ to get its upper limit, as shown in the lower panel of Figure \ref{fig:Frobs}.

In Table \ref{table:Tc}, we summarize the constraints on $T_c$ with the CHIME data point $t$ and $F_{\rm r,obs}$ values for the assumed range of $k$. For each parameter set, combining the lower and upper limits together, one can constrain $T_c$ to a small range. However, since some parameters of the repeaters' population have not been determined yet, the range of the possible $T_c$ would be very large, which is $0.1d<T_c< \infty $. Considering those constraints on $T_c$, we could predict the observed fraction of repeaters in the future with a CHIME-like telescope array, with a duty cycle as assumed in this paper. Our predictions are shown in Table \ref{table:Fr}. The bold fonts represent the cases in which the peak should have appeared, whereas the thin fonts represent the cases in which the peak may have not appeared yet. We find that for all the parameter sets, there would not be a significant increase of $F_{\rm r,obs}$ with $3$-year, $10$-year or even $30$-year observations. In most cases, the peak would be observed in $30$ years. The peak fraction of repeaters would not be more than $0.04$ in all cases.

\begin{table*}
\caption{Constraints on $T_c$ with current data.}
\centering
\begin{tabular}{|c|c|c|c|c|c|}
\hline 
\backslashbox[40mm]{$q$}{$r_{\rm 0,min}$ ({\rm $hr^{-1}$})} & $10^{-3.5}$ & $10^{-4.5}$ & $10^{-5.5}$ & $10^{-6.5}$ & $10^{-10.5}$ \\
\hline
\multirow{2}{*}{1.3} & \multirow{2}{*}{$T_c<0.1 d $}& \multirow{2}{*}{$T_c<0.4 d$} & \multirow{2}{*}{$0.8d <T_c < 1.3d$} & \multirow{2}{*}{$T_c<2.0d$} &  \multirow{2}{*}{$T_c < 35d$} \\
&  &  &  & & \\
\hline
\multirow{2}{*}{1.6} & \multirow{2}{*}{$0.1d<T_c<1.0d$} & \multirow{2}{*}{$1.2d<T_c<4.6d$} & \multirow{2}{*}{$5d<T_c<16.5d$} & \multirow{2}{*}{$15d<T_c<\infty$}& \multirow{2}{*}{$T_c \rightarrow \infty $}  \\
&  &  &  & & \\
\hline 
\multirow{2}{*}{1.8} & \multirow{2}{*}{$T_c<2.3d$} & \multirow{2}{*}{$T_c<13d$} & \multirow{2}{*}{$15d<T_c<\infty$} & \multirow{2}{*}{$3000d<T_c< \infty$} & \multirow{2}{*}{$--$}  \\
& & & & & \\
\hline 
\end{tabular}
\label{table:Tc}
\end{table*}

\begin{table*}
\caption{Predictions for future observation}
\centering
\begin{tabular}{|c|c|c|c|c|c|}
\hline 
\multirow{2}{*}{\backslashbox[40mm]{$t$}{$r_0$ }} & $r_{\rm 0,min}=10^{-3.5}$ & $r_{\rm 0,min}=10^{-4.5}$ & \multicolumn{3}{c|}{$r_{\rm 0,min}=10^{-5.5}$}   \\
\cline{2-6}
&$q=1.6$ &$q=1.6$ & $q=1.3$ & $q=1.6$ & $q=1.8$ \\
\hline
\multirow{2}{*}{$3{\rm yr}$} & \multirow{2}{*}{$F_{\rm r,obs}<0.04$} & \multirow{2}{*}{$F_{\rm r,obs}<0.03$} & \multirow{2}{*}{$F_{\rm r,obs}<0.02$} & \multirow{2}{*}{${ F_{\rm r,obs}<0.02}$} & \multirow{2}{*}{$ F_{\rm r,obs}<0.013$} \\
& & & & &  \\
\hline
\multirow{2}{*}{$10{\rm yr}$} & \multirow{2}{*}{$F_{\rm r,obs}<0.04$} &  \multirow{2}{*}{$F_{\rm r,obs}<0.03$} & \multirow{2}{*}{${\bf F_{\rm r,obs,p}<0.04}$}&  \multirow{2}{*}{${\bf F_{\rm r,obs,p}<0.02}$} & \multirow{2}{*}{$F_{\rm r,obs}<0.013$}  \\

& & & & &  \\

\hline
\multirow{2}{*}{$30{\rm yr}$} & \multirow{2}{*}{ ${\bf F_{\rm r,obs,p}<0.04}$} & \multirow{2}{*}{${\bf F_{\rm r,obs,p}<0.03}$} & \multirow{2}{*}{${\bf F_{\rm r,obs,p}<0.04}$} & \multirow{2}{*}{${\bf F_{\rm r,obs,p}<0.02}$} & \multirow{2}{*}{$F_{\rm r,obs}<0.015$} \\
& & & & &  \\
\hline 
\end{tabular} \\
\label{table:Fr}
\end{table*}

\section{conclusions and discussion}\label{sec:summary}

We have introduced a parameter $T_c$ ($T_c^{~\prime}$ for evolving repeaters) (defined in Equation \ref{eq:Tc}, or Equation \ref{eq:Tcp} for the evolving case) to describe the real fraction of repeating FRB sources in the entire universe. A smaller $T_c$ means a larger fraction of genuinely non-repeating sources per unit time. The ansatz that all FRBs repeat corresponds to 
$T_c \rightarrow \infty$. Consider that not all repeating sources can be recognized, we performed a set of Monte Carlo simulations to investigate how the observed repeating source fraction  $F_{\rm r,obs}$ is related to the true fraction ($T_c$ as a proxy).

Assuming that all repeaters have the same intrinsic repeating rate, we found the following:
\begin{itemize}
    \item If all repeaters are non-evolving, once a finite $T_c$ is introduced, $F_{\rm r,obs}$ would not always increase with time, but instead shows a turnover after reaching a peak value $F_{\rm r,obs,p}$ at a peak time $T_p$.  The expected turnover point is highly dependent on $T_c$, with a larger $T_c$ corresponding to a higher $F_{\rm r,obs,p}$ and a later $T_p$.
    
    \item If the evolution of repeaters is considered, a turnover point is expected only if $T_p$ is smaller than their characteristic lifetime $T_l$. When the observational timescale is longer than $T_l$, $F_{\rm r,obs}$ would roughly be a constant. The curve of $F_{\rm r,obs}$ would be flattened even if $T_c^{~\prime}\rightarrow \infty$.

\end{itemize}

The $(T_p, F_{\rm r,obs,p})$ pair  also depends on the repeating rate distribution of the repeaters as well as time distribution function of the bursts (the Weibull parameter $k$). Assume the intrinsic repeating rate $r_0$ satisfies a power-low distribution with an index $q$ in the range of [$r_{\rm 0,rmin}$, $r_{\rm 0, rmax}$] and fix $r_{\rm 0,max}$ (which could be measured from known repeaters). We found the following:
\begin{itemize}
    \item In the non-evolving repeater approximation, a higher $q$ and a lower $r_{\rm 0,min}$ would both result in a distribution with more inactive repeating sources, which would lead to a lower $F_{\rm r,obs,p}$ and an earlier $T_p$. With other parameters fixed, the $k$ parameter would dramatically influence $T_p$ but only slightly change $F_{\rm r,obs,p}$.
    \item Assume that the distribution of $r_0$ at a specific time is introduced by the evolution of $r_0$ with time and that the lifetime of the repeaters which is longer than the observational timescale. The trend on how different parameters influence $T_p$ and $F_{\rm r,obs,p}$ remains the same as that in non-evolving approximation case.
\end{itemize}

Available CHIME observations give $F_{\rm r,obs}\sim 0.013$ at  $t \sim 400$ d. One may regard this as a lower limit of $F_{\rm,obs,p}$ and $T_p$, so that the data can already place a lower limit on $T_c$, i.e. $T_c > 0.1$ d with reasonable parameters. In the future, if a higher value of $F_{\rm r,obs}$ is observed, a more stringent lower limit on $T_c$ can be obtained. The theoretical evolution curve of $F_{r,obs}$ should always pass through the current data point, thus an upper limit on $T_c$ could be also obtained albeit with a large uncertainty. We predict that the observed fraction of repeaters would remain smaller than $0.04$ with a $30$-year observations with CHIME or similar telescope arrays. If a peak fraction smaller than $0.04$ is actually observed in the near future, the ansatz that all FRB sources repeat would be disfavored.

All the conclusions drawn in this paper are based on the assumption that all the repeaters have the same $r_0$ or $r_0$ follows a power-law distribution. So far there is no direct measurement of the $r_0$ distribution. However, such a distribution can be readily measured when more repeating FRBs are closely monitored. If in the future the $r_0$ distribution is proven not a power law, some of our conclusions need to be re-investigated.

\acknowledgments
SA and BZ acknowledges the Top Tier Doctoral Graduate Research Assistantship (TTDGRA) at University of Nevada, Las Vegas for support. HG acknowledges the National Natural Science Foundation of China (NSFC) under Grant No. 11722324, 11690024 and the Fundamental Research Funds for the Central Universities for support. BZ thanks Xuelei Chen for a stimulative conversation.

\appendix

\section{Weibull distribution}
The distribution of the time interval of two adjacent bursts in a repeating source could be describe by a Wellbull function, which reads 
\begin{eqnarray}
{\cal W}(\delta | k,r)=k \delta^{-1}\left[\delta r \Gamma(1+1/k)\right]^k e^{-\left[\delta r \Gamma(1+1/k)\right]^k},
\label{eq:Wb}
\end{eqnarray}
where $r$ represents the mean repeating rate, $k$ is the shape parameter, and $\Gamma(x)$ stands for the Gamma function. 

When $k=1$, the distribution is reduced to the exponential (i.e. Poisson) distribution. In this case, Equation \ref{eq:Wb} is simplified as 
\begin{eqnarray}
{\cal W}(\delta | r)=re^{-r\delta}.
\end{eqnarray}
The mean interval time is
\begin{eqnarray}
\left<\delta \right>=\int_{0}^{\infty}\delta {\cal W}(\delta |r) d\delta=1/r.
\end{eqnarray}
The variance of $\delta$ could be calculated as
\begin{eqnarray}
D(\delta)&=& \left<\delta^2\right>- \left<\delta \right>^2 \nonumber \\
&=&\int_{0}^{\infty}\delta^2{\cal W}(\delta |r) d\delta-{1 \over r^2} \nonumber \\
&=&{2 \over r^2 }-{1\over r^2} \nonumber \\
&=&{1 \over r^2}.
\end{eqnarray}

Assume that an observation starts at time $t_s$ after the first burst and the waiting time until the next burst appears is $t_w$. The probability that an observer would wait for at least a period of $t_1$ could be written as
\begin{eqnarray}
P(t_w>t_1)=P(\delta>t_s+t_1 | \delta>t_s)&=&{e^{-r(t_1+t_s)} \over e^{-rt_s}} \nonumber \\
&=&e^{-rt_1}.
\end{eqnarray}
Similarly, the probability of having a burst in $t_1$ would be
\begin{eqnarray}
P(t_w<t_1)=1-e^{-rt_1},
\end{eqnarray}
which is independent of $t_s$. The probability density function is 
\begin{eqnarray}
{\cal W}(t_w | r)=re^{-rt_w},
\end{eqnarray}
which is the same as that of the time interval $\delta$. Therefore, the mean waiting time is the same as the true mean interval time between two adjacent bursts.

When $k \neq 1$, one can similarly calculate the mean time interval as
\begin{eqnarray}
\left<\delta \right>&=&\int_{0}^{\infty} \delta{\cal W}(\delta |r)d\delta\nonumber \\
&=& \int_0^{\infty} k[\delta r \Gamma(1+1/k)]^k e^{-[\delta r \Gamma(1+1/k)]^k} d\delta \nonumber \\
&=& {1 \over r\Gamma(1+1/k)} \int_0^{\infty} x^{1/k} e^{-x}dx,
\end{eqnarray}
where $x=[\delta r \Gamma(1+1/k)]^k$. Considering that
$\Gamma(1+z)=\int_0^{\infty}x^z e^{-x} dx$, one has
\begin{eqnarray}
\left<\delta \right>={1 \over r\Gamma(1+1/k)} \times \Gamma(1+1/k)={1\over r}.
\end{eqnarray}
The mean time interval does not change with $k$.

We can calculate the variance of the Weibull distribution to study the clustering effect introduced by the shape parameter $k$. The variance is
\begin{eqnarray}
D(\delta)&=& \left<\delta^2 \right>- \left<\delta \right>^2 \nonumber \\
&=& \int_0^{\infty} \delta^2 {\cal W}(\delta |r)d\delta - {1 \over r^2} \nonumber \\
&=&{1 \over [r\Gamma(1+1/k)]^2}\int_0^{\infty} x^{2/k} e^{-x} dx - {1 \over r^2} \nonumber \\
&=&{1 \over r^2} \left[{\Gamma(1+2/k) \over \Gamma(1+1/k)^2}-1\right].
\end{eqnarray}

\begin{figure*}[ht!]
\centering
\resizebox{90mm}{!}{\includegraphics[]{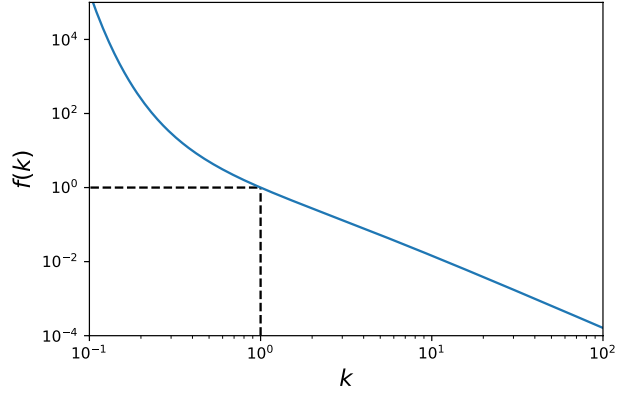}}
\caption{The correcting factor $f(k)$ of the variance of Weibull distribution compared with exponential distribution as a function of shape parameter $k$.} 
\label{fig:shape}
\end{figure*}

Compared with exponential distribution, the variance of Weibull distribution is corrected by a factor 
\begin{eqnarray}
f(k)={\Gamma(1+2/k) \over \Gamma(1+1/k)^2}-1
\end{eqnarray}
as a function of $k$, which is shown in Figure \ref{fig:shape}.

When $k<1$, the variance of Weibull distribution is larger than ${1/r^2}$, which means that both longer and shorter time intervals have more chance to appear. That would lead to the case that some bursts are closer to each other and some others are more separated from each other. Hence we tend to detect ``clusters'' of bursts. When $k>1$, the variance of Weibull distribution would be smaller than ${1/r^2}$, which means that the time intervals between bursts tend to be the same. In this case, the repeating burst would appear to be more ``periodic''.

In addition, when $k \neq 1$, the waiting time distribution is not the same as the time interval distribution.


\begin{thebibliography}{}
\expandafter\ifx\csname natexlab\endcsname\relax\def\natexlab#1{#1}\fi
\providecommand{\url}[1]{\href{#1}{#1}}
\providecommand{\dodoi}[1]{doi:~\href{http://doi.org/#1}{\nolinkurl{#1}}}
\providecommand{\doeprint}[1]{\href{http://ascl.net/#1}{\nolinkurl{http://ascl.net/#1}}}
\providecommand{\doarXiv}[1]{\href{https://arxiv.org/abs/#1}{\nolinkurl{https://arxiv.org/abs/#1}}}

\bibitem[{{Bochenek} {et~al.}(2020){Bochenek}, {Ravi}, {Belov}, {Hallinan},
  {Kocz}, {Kulkarni}, \& {McKenna}}]{Bochenek20}
{Bochenek}, C.~D., {Ravi}, V., {Belov}, K.~V., {et~al.} 2020, \nat, 587, 59,
  \dodoi{10.1038/s41586-020-2872-x}

\bibitem[{{Caleb} {et~al.}(2019){Caleb}, {Stappers}, {Rajwade}, \&
  {Flynn}}]{caleb19}
{Caleb}, M., {Stappers}, B.~W., {Rajwade}, K., \& {Flynn}, C. 2019, \mnras,
  484, 5500, \dodoi{10.1093/mnras/stz386}

\bibitem[{{CHIME/FRB Collaboration} {et~al.}(2019{\natexlab{a}}){CHIME/FRB
  Collaboration}, {Amiri}, {Bandura}, {Bhardwaj}, {Boubel}, {Boyce}, {Boyle},
  {. Brar}, {Burhanpurkar}, {Cassanelli}, {Chawla}, {Cliche}, {Cubranic},
  {Deng}, {Denman}, {Dobbs}, {Fandino}, {Fonseca}, {Gaensler}, {Gilbert},
  {Gill}, {Giri}, {Good}, {Halpern}, {Hanna}, {Hill}, {Hinshaw}, {H{\"o}fer},
  {Josephy}, {Kaspi}, {Landecker}, {Lang}, {Lin}, {Masui}, {Mckinven},
  {Mena-Parra}, {Merryfield}, {Michilli}, {Milutinovic}, {Moatti}, {Naidu},
  {Newburgh}, {Ng}, {Patel}, {Pen}, {Pinsonneault-Marotte}, {Pleunis},
  {Rafiei-Ravandi}, {Rahman}, {Ransom}, {Renard}, {Scholz}, {Shaw}, {Siegel},
  {Smith}, {Stairs}, {Tendulkar}, {Tretyakov}, {Vanderlinde}, \&
  {Yadav}}]{chime19a}
{CHIME/FRB Collaboration}, {Amiri}, M., {Bandura}, K., {et~al.}
  2019{\natexlab{a}}, \nat, 566, 235, \dodoi{10.1038/s41586-018-0864-x}

\bibitem[{{CHIME/FRB Collaboration} {et~al.}(2019{\natexlab{b}}){CHIME/FRB
  Collaboration}, {Andersen}, {Bandura}, {Bhardwaj}, {Boubel}, {Boyce},
  {Boyle}, {Brar}, {Cassanelli}, {Chawla}, {Cubranic}, {Deng}, {Dobbs},
  {Fandino}, {Fonseca}, {Gaensler}, {Gilbert}, {Giri}, {Good}, {Halpern},
  {Hill}, {Hinshaw}, {H{\"o}fer}, {Josephy}, {Kaspi}, {Kothes}, {Landecker},
  {Lang}, {Li}, {Lin}, {Masui}, {Mena-Parra}, {Merryfield}, {Mckinven},
  {Michilli}, {Milutinovic}, {Naidu}, {Newburgh}, {Ng}, {Patel}, {Pen},
  {Pinsonneault-Marotte}, {Pleunis}, {Rafiei-Ravandi}, {Rahman}, {Ransom},
  {Renard}, {Scholz}, {Siegel}, {Singh}, {Smith}, {Stairs}, {Tendulkar},
  {Tretyakov}, {Vanderlinde}, {Yadav}, \& {Zwaniga}}]{chime19b}
{CHIME/FRB Collaboration}, {Andersen}, B.~C., {Bandura}, K., {et~al.}
  2019{\natexlab{b}}, \apjl, 885, L24, \dodoi{10.3847/2041-8213/ab4a80}

\bibitem[{{Cordes} \& {Chatterjee}(2019)}]{cordes19}
{Cordes}, J.~M., \& {Chatterjee}, S. 2019, \araa, 57, 417,
  \dodoi{10.1146/annurev-astro-091918-104501}

\bibitem[{{Fonseca} {et~al.}(2020){Fonseca}, {Andersen}, {Bhardwaj}, {Chawla},
  {Good}, {Josephy}, {Kaspi}, {Masui}, {Mckinven}, {Michilli}, {Pleunis},
  {Shin}, {Tendulkar}, {Bandura}, {Boyle}, {Brar}, {Cassanelli}, {Cubranic},
  {Dobbs}, {Dong}, {Gaensler}, {Hinshaw}, {Land ecker}, {Leung}, {Li}, {Lin},
  {Mena-Parra}, {Merryfield}, {Naidu}, {Ng}, {Patel}, {Pen}, {Rafiei-Ravandi},
  {Rahman}, {Ransom}, {Scholz}, {Smith}, {Stairs}, {Vanderlinde}, {Yadav}, \&
  {Zwaniga}}]{fonseca20}
{Fonseca}, E., {Andersen}, B.~C., {Bhardwaj}, M., {et~al.} 2020, \apjl, 891,
  L6, \dodoi{10.3847/2041-8213/ab7208}

\bibitem[{{James}(2019)}]{james19}
{James}, C.~W. 2019, \mnras, 486, 5934, \dodoi{10.1093/mnras/stz1224}

\bibitem[{{Law} {et~al.}(2017){Law}, {Abruzzo}, {Bassa}, {Bower},
  {Burke-Spolaor}, {Butler}, {Cantwell}, {Carey}, {Chatterjee}, {Cordes},
  {Demorest}, {Dowell}, {Fender}, {Gourdji}, {Grainge}, {Hessels}, {Hickish},
  {Kaspi}, {Lazio}, {McLaughlin}, {Michilli}, {Mooley}, {Perrott}, {Ransom},
  {Razavi-Ghods}, {Rupen}, {Scaife}, {Scott}, {Scholz}, {Seymour}, {Spitler},
  {Stovall}, {Tendulkar}, {Titterington}, {Wharton}, \& {Williams}}]{law17}
{Law}, C.~J., {Abruzzo}, M.~W., {Bassa}, C.~G., {et~al.} 2017, \apj, 850, 76,
  \dodoi{10.3847/1538-4357/aa9700}

\bibitem[{{Lorimer} {et~al.}(2007){Lorimer}, {Bailes}, {McLaughlin},
  {Narkevic}, \& {Crawford}}]{lorimer07}
{Lorimer}, D.~R., {Bailes}, M., {McLaughlin}, M.~A., {Narkevic}, D.~J., \&
  {Crawford}, F. 2007, Science, 318, 777, \dodoi{10.1126/science.1147532}

\bibitem[{{Lu} \& {Piro}(2019)}]{lu19}
{Lu}, W., \& {Piro}, A.~L. 2019, \apj, 883, 40,
  \dodoi{10.3847/1538-4357/ab3796}

\bibitem[{{Lu} {et~al.}(2020){Lu}, {Piro}, \& {Waxman}}]{lu20}
{Lu}, W., {Piro}, A.~L., \& {Waxman}, E. 2020, \mnras,
  \dodoi{10.1093/mnras/staa2397}

\bibitem[{{Luo} {et~al.}(2018){Luo}, {Lee}, {Lorimer}, \& {Zhang}}]{luo18}
{Luo}, R., {Lee}, K., {Lorimer}, D.~R., \& {Zhang}, B. 2018, \mnras, 481, 2320,
  \dodoi{10.1093/mnras/sty2364}

\bibitem[{{Luo} {et~al.}(2020){Luo}, {Men}, {Lee}, {Wang}, {Lorimer}, \&
  {Zhang}}]{luo20}
{Luo}, R., {Men}, Y., {Lee}, K., {et~al.} 2020, \mnras, 494, 665,
  \dodoi{10.1093/mnras/staa704}

\bibitem[{{Oppermann} {et~al.}(2018){Oppermann}, {Yu}, \& {Pen}}]{oppermann18}
{Oppermann}, N., {Yu}, H.-R., \& {Pen}, U.-L. 2018, \mnras, 475, 5109,
  \dodoi{10.1093/mnras/sty004}

\bibitem[{{Palaniswamy} {et~al.}(2018){Palaniswamy}, {Li}, \&
  {Zhang}}]{palaniswamy18}
{Palaniswamy}, D., {Li}, Y., \& {Zhang}, B. 2018, \apjl, 854, L12,
  \dodoi{10.3847/2041-8213/aaaa63}

\bibitem[{{Petroff} {et~al.}(2019){Petroff}, {Hessels}, \&
  {Lorimer}}]{petroff19}
{Petroff}, E., {Hessels}, J.~W.~T., \& {Lorimer}, D.~R. 2019, \aapr, 27, 4,
  \dodoi{10.1007/s00159-019-0116-6}

\bibitem[{{Petroff} {et~al.}(2015){Petroff}, {Johnston}, {Keane}, {van
  Straten}, {Bailes}, {Barr}, {Barsdell}, {Burke-Spolaor}, {Caleb}, {Champion},
  {Flynn}, {Jameson}, {Kramer}, {Ng}, {Possenti}, \& {Stappers}}]{petroff15}
{Petroff}, E., {Johnston}, S., {Keane}, E.~F., {et~al.} 2015, \mnras, 454, 457,
  \dodoi{10.1093/mnras/stv1953}

\bibitem[{{Ravi}(2019)}]{ravi19}
{Ravi}, V. 2019, Nature Astronomy, 3, 928, \dodoi{10.1038/s41550-019-0831-y}

\bibitem[{{Spitler} {et~al.}(2016){Spitler}, {Scholz}, {Hessels}, {Bogdanov},
  {Brazier}, {Camilo}, {Chatterjee}, {Cordes}, {Crawford}, {Deneva}, {Ferdman},
  {Freire}, {Kaspi}, {Lazarus}, {Lynch}, {Madsen}, {McLaughlin}, {Patel},
  {Ransom}, {Seymour}, {Stairs}, {Stappers}, {van Leeuwen}, \&
  {Zhu}}]{spitler16}
{Spitler}, L.~G., {Scholz}, P., {Hessels}, J.~W.~T., {et~al.} 2016, \nat, 531,
  202, \dodoi{10.1038/nature17168}

\bibitem[{{The CHIME/FRB Collaboration} {et~al.}(2020){The CHIME/FRB
  Collaboration}, {:}, {Andersen}, {Band ura}, {Bhardwaj}, {Bij}, {Boyce},
  {Boyle}, {Brar}, {Cassanelli}, {Chawla}, {Chen}, {Cliche}, {Cook},
  {Cubranic}, {Curtin}, {Denman}, {Dobbs}, {Dong}, {Fandino}, {Fonseca},
  {Gaensler}, {Giri}, {Good}, {Halpern}, {Hill}, {Hinshaw}, {H{\"o}fer},
  {Josephy}, {Kania}, {Kaspi}, {Landecker}, {Leung}, {Li}, {Lin}, {Masui},
  {Mckinven}, {Mena-Parra}, {Merryfield}, {Meyers}, {Michilli}, {Milutinovic},
  {Mirhosseini}, {M{\"u}nchmeyer}, {Naidu}, {Newburgh}, {Ng}, {Patel}, {Pen},
  {Pinsonneault-Marotte}, {Pleunis}, {Quine}, {Rafiei-Ravandi}, {Rahman},
  {Ransom}, {Renard}, {Sanghavi}, {Scholz}, {Shaw}, {Shin}, {Siegel}, {Singh},
  {Smegal}, {Smith}, {Stairs}, {Tan}, {Tendulkar}, {Tretyakov}, {Vanderlinde},
  {Wang}, {Wulf}, \& {Zwaniga}}]{CHIME20}
{The CHIME/FRB Collaboration}, {:}, {Andersen}, B.~C., {et~al.} 2020, arXiv
  e-prints, arXiv:2005.10324.
\newblock \doarXiv{2005.10324}

\bibitem[{{Thornton} {et~al.}(2013){Thornton}, {Stappers}, {Bailes},
  {Barsdell}, {Bates}, {Bhat}, {Burgay}, {Burke-Spolaor}, {Champion}, {Coster},
  {D'Amico}, {Jameson}, {Johnston}, {Keith}, {Kramer}, {Levin}, {Milia}, {Ng},
  {Possenti}, \& {van Straten}}]{thornton13}
{Thornton}, D., {Stappers}, B., {Bailes}, M., {et~al.} 2013, Science, 341, 53,
  \dodoi{10.1126/science.1236789}

\bibitem[{{Y{\"u}ksel} {et~al.}(2008){Y{\"u}ksel}, {Kistler}, {Beacom}, \&
  {Hopkins}}]{yuksel08}
{Y{\"u}ksel}, H., {Kistler}, M.~D., {Beacom}, J.~F., \& {Hopkins}, A.~M. 2008,
  \apjl, 683, L5, \dodoi{10.1086/591449}

\end{thebibliography}
\end{document}